# In-plane dielectric constant and conductivity of confined water

R. Wang[1,2,+], M. Souilamas[1,2,+], A. Esfandiar[1,2], R. Fabregas[1], S. Benaglia[1,2], H. Nevison-Andrews[1,2], Q. Yang[1,2], J. Normansell[1,2], P. Ares[1,2], G. Ferrari[3], A. Principi[1], A. K. Geim[1,2*], L. Fumagalli[1,2*]

[1]Department of Physics & Astronomy University of Manchester, Manchester, M13 9PL, UK
[2]National Graphene Institute, University of Manchester, Manchester, M13 9PL, UK
[3]Department of Physics, Politecnico di Milano, Via Colombo 81, 20133 Milano, Italy
[+] These authors contributed equally to this work
*Corresponding authors: L.F. (laura.fumagalli@manchester.ac.uk) and A.K.G. (geim@manchester.ac.uk)

**Water is essential for almost every aspect of life on our planet and, unsurprisingly, its properties have been studied in great detail[1]. However, disproportionately little remains known about the electrical properties of interfacial and strongly confined water[2,3] where its structure deviates from that of bulk water, becoming distinctly layered[4,5]. The structural change is expected to affect water's conductivity and particularly its polarizability, which in turn modifies intermolecular forces that play a crucial role in many physical and chemical processes[6-9]. Here we use scanning dielectric microscopy[10] to probe the in-plane electrical properties of water confined between atomically flat surfaces separated by distances down to 1 nm. For confinement exceeding a few nm, water exhibits an in-plane dielectric constant close to that of bulk water and its proton conductivity is notably enhanced, gradually increasing with decreasing water thickness. This trend abruptly changes when the confined water becomes only a few molecules thick. Its in-plane dielectric constant reaches giant, ferroelectric-like values of about 1,000 whereas the conductivity peaks at a few S·m$^{-1}$, close to values characteristic of superionic liquids. We attribute the enhancement to strongly disordered hydrogen bonding induced by the few-layer confinement, which facilitates both easier in-plane polarization of molecular dipoles and faster proton exchange. This insight into the electrical properties of nanoconfined water is important for understanding many phenomena that occur at aqueous interfaces and in nanoscale pores.**

In the bulk and at room temperature (RT), water exhibits exceptionally high dielectric constant ($\varepsilon_{bulk} \approx 80$) and high (for a wide-bandgap insulator) electrical conductivity ($\sigma_{bulk} \approx 10^{-5}$ S·m$^{-1}$)[1]. Both characteristics are inherently connected with the ability of water molecules to form hydrogen bonds[11-13] and are key to water's main properties. Among them is its remarkable ability to dissolve more substances than any other liquid[7,8], which stems from water's high $\varepsilon_{bulk}$ that efficiently suppresses Coulomb interactions between solutes. The strong dielectric screening is also critical in biochemical processes responsible for life[7] (e.g., proteins' folding and assembly, their interaction with nucleic acids, and ion transport across cell membranes). Importantly, a multitude of phenomena involving water (including solvation) occurs at solid interfaces where liquid water exhibits a distinct layered structure[4,5]. Within these few layers, the hydrogen-bond network is greatly altered as compared to that of bulk water, no longer following the ice rules[14]. Accordingly, the electrical properties of water near surfaces and inside nanoscale cavities are expected to be different from those of bulk water[2,3]. These differences have been the subject of intense research. In particular, proton conductivity of near-surface water has been reported to become much larger than $\sigma_{bulk}$, although the magnitude of this enhancement and the underlying mechanism remain debated[15-19]. Meanwhile, it has proven difficult to assess the dielectric properties of interfacial and nanoconfined water[20,21]. Only recently[22] it has been shown that, in the direction perpendicular to surfaces, water within a few near-surface layers becomes nonpolarizable, exhibiting the dielectric constant $\varepsilon_\perp \approx 2$, in agreement with many theoretical predictions[23-26]. However, the in-plane dielectric constant $\varepsilon_{//}$ of interfacial water remains essentially unknown, because $\varepsilon_{//}$ is even more challenging to measure than $\varepsilon_\perp$ for the lack of suitable experimental techniques. The in-plane polarizability of near-surface water is also poorly understood theoretically and, depending on assumptions and confinement strength, $\varepsilon_{//}$ was predicted to attain values from giant to bulk-like to exceptionally small[27-31].

## Scanning dielectric microscopy on water-filled nanochannels

To probe the in-plane response of nanoconfined water, we have employed scanning dielectric microscopy (SDM)[10]. Our setup is schematically shown in Fig. 1a and detailed in Methods. Briefly, nanoscale channels of various heights $h$ and having the width of $\sim$ 200 nm were fabricated by van der Waals assembly of atomically flat monocrystals of hexagonal boron nitride (hBN) and graphite (for fabrication details, see Methods). An AC voltage was applied to an AFM tip, and its mechanical response to the electrostatic force translated into local electrical impedance. By scanning the tip across the channels, the dielectric response of water inside the nanochannels was measured and compared with that of nearby insulating spacers made of hBN with the known dielectric constant ($\varepsilon_{hBN} \approx 4$). The approach is conceptually similar to that used to study $\varepsilon_\perp$ of nanoconfined water[22] but the previous setup was insensitive to $\varepsilon_{//}$ because the electric field $E$ was applied perpendicular to the channels. To measure $\varepsilon_{//}$, we have introduced two main changes. First, the ground electrode (graphite crystal in Fig. 1a; Extended Data Fig. 1) has been moved away from the probed water layer by adding a relatively thick hBN crystal to the assembly, which creates an in-plane field component $E_{//}$ inside the water (Fig. 1a). Second, we have expanded the measurement bandwidth by $\sim$ 5 orders of magnitude from the standard kHz frequencies into the GHz regime, which is challenging but essential for measuring $\varepsilon_{//}$ in the presence of unexpectedly high $\sigma_{//}$ found for few-layer (quasi-2D) water (see below).

Representative SDM images from our relatively large (reference) channels with $h \approx 30$ nm (Fig. 1b) are shown in Figs. 1c and 1d before and after filling them with deionized water, respectively (also see Extended Data Fig. 2). The dielectric contrast between empty channels ($\varepsilon = 1$) and hBN spacers appears negative as expected, reflecting the relatively larger dielectric constant of hBN. After the channels were filled, the contrast changed and became strongly positive, confirming the presence of water inside the channels and its strong dielectric response. With increasing the measurement frequency, $f$, the contrast for empty channels did not change, again as expected (Fig. 1c). For water-filled channels, it remained positive with increasing $f$ from kHz to MHz and GHz but weakened (Fig. 1d). For each $f$, we then plotted the average contrast of water with respect to hBN (Fig. 1e) and, by repeating the imaging at many different $f$ (see detailed measurement and calibration procedures in Methods; Extended Data Fig. 3), we obtained dielectric dispersion curves such as those shown in Fig. 2a (see the blue symbols for $h \approx 30$ nm).

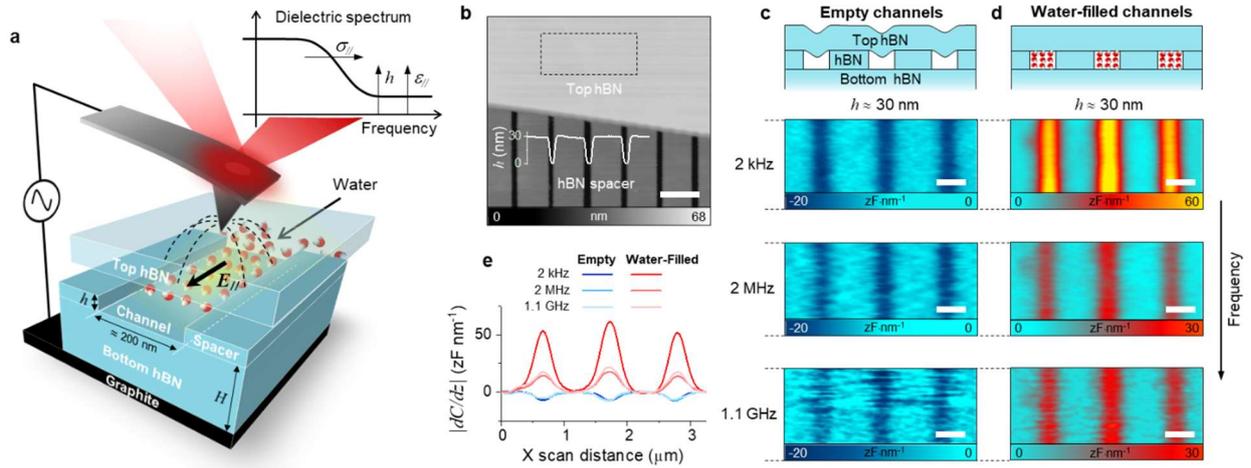

**Fig. 1| Broadband dielectric imaging of nanoconfined water.** (**a**) Schematics of the experimental setup and our devices. The hBN top, bottom and spacer layers are shown in light blue and the graphite ground electrode in black. The water is filled from a backside inlet (Extended Data Fig. 1d). Oscillating voltage ($f$ from 0.1 kHz to 1.1 GHz) is applied to the AFM tip, and the measured local impedance is converted into dielectric spectra such as that schematically shown in the inset. Relatively thick bottom hBN ($H \approx 50–200$ nm) is used to create in-plane field $E_{//}$ inside water channels. (**b**) AFM micrograph of a device with $h \approx 30$ nm with the inset showing a trace over the exposed spacers. (**c** and **d**) Dielectric images at characteristic $f$ of 2 kHz, 2 MHz and 1.1 GHz for the same three channels before and after filling them with water, respectively. The images were taken at 295 ± 1 K from the area outlined in (b) by the dashed rectangle. The top panels in (c) and (d) sketch the channels' cross-section (not to scale). The slight sagging of the top hBN into empty channels disappears after filling them with water. The sagging was monitored by AFM to ensure that the channels were fully filled[22]. (**e**) Dielectric profiles (averaged over 10 lines) from (c,d); color coded. $|dC/dz|$ is defined in Methods. Scale bars: (b) 1 μm, (c-d) 500 nm.

Although the experimental curve resembles the Debye dispersion, the orientational relaxation of water dipoles responsible for the Debye behavior of bulk water plays little role at $f < 10$ GHz, and the dispersion found at our relatively low $f$ is caused by conductive losses (Methods). Indeed, at high $f$, the dielectric response is determined by water's $\varepsilon_{//}$ and is independent of $\sigma_{//}$, which yields the high-$f$ plateau with its strength increasing with increasing $\varepsilon_{//}$ and $h$. Water's $\sigma_{//}$ starts contributing at $f$ below the conductivity relaxation frequency $f_r = \sigma_{//}/2\pi\varepsilon_0\varepsilon_{//}$, giving rise to a stronger dielectric contrast. The conductivity contribution is cut off at $f_c = \alpha\sigma_{//}/2\pi\varepsilon_0$ by capacitive coupling between the water and the bottom graphite as well as the AFM tip, which results in the low-$f$ plateau. The cut-off frequency $f_c$ increases proportionally to $\sigma_{//}$ and is independent of $\varepsilon_{//}$, where $\alpha$ is the factor that depends on $h$ and the measurement geometry. The inset of Fig. 1a summarizes the tendencies expected for the dielectric response with changing $h$, $\varepsilon_{//}$ and $\sigma_{//}$. Further details about SDM measurements are provided in Methods (Extended Data Figs. 4-10).

**Dielectric dispersion of nanoconfined water**

In addition to the reference spectrum from the 30-nm channels, Fig. 2a shows the spectra obtained for ultrathin water in channels with other representative heights $h \approx 5$ and 1.5 nm (cyan and red) whereas Fig. 2b provides examples of their SDM imaging (also see Extended Data Fig. 4). One can see two clear trends in how the dielectric response of water evolves with decreasing $h$. First, the transition between low- and high-$f$ plateaus shifts to higher frequencies, from kHz to MHz. This is also apparent from the images of Fig. 2b where changes in the contrast occur at higher $f$ for the thinner channel. The behavior unambiguously shows that water becomes more conductive under strong confinement. This is somewhat expected because water is known to interact with hBN walls, which results in near-surface dissociation of water molecules and higher proton conductivity[17,32]. The strongly enhanced $\sigma$ is also consistent with the previous results for water inside hBN nanochannels and nanoporous media[33,34]. Second, the high-$f$ dielectric response does not gradually diminish with decreasing $h$ as expected if $\varepsilon$ of nanoconfined water were to remain the same as of bulk water (cf. inset of Fig. 1a). Instead, the high-$f$ plateaus for $h \approx 1.5$ and 5 nm in Fig. 2a remain of similar height. This observation qualitatively implies that water's $\varepsilon$ notably increases under strong confinement. The positive contrast at high $f$ for $h \approx 1.5$ nm is particularly striking if compared with the behavior found by SDM without a parallel component of $E$. In the latter case[22], the dielectric contrast reversed from positive to negative at $h \approx 3$ nm because of diminishing $\varepsilon_\perp$. This does not happen for the in-plane measurement geometry even for our smallest $h \approx 1$ nm, and again implies large $\varepsilon_{//}$ for water inside atomic-scale channels.

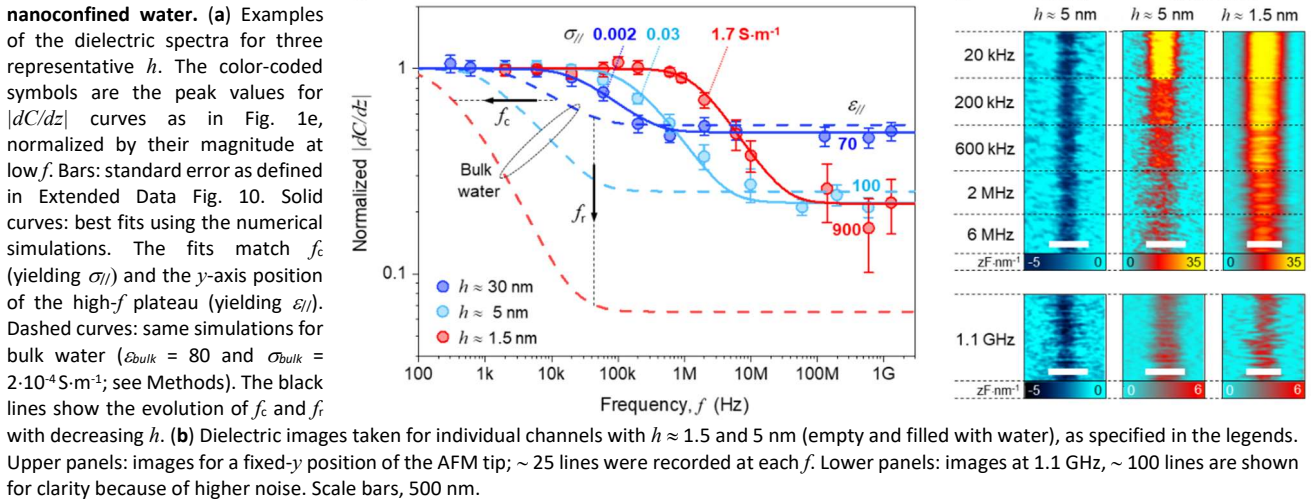

**Fig. 2| Dielectric dispersion of nanoconfined water.** (**a**) Examples of the dielectric spectra for three representative $h$. The color-coded symbols are the peak values for $|dC/dz|$ curves as in Fig. 1e, normalized by their magnitude at low $f$. Bars: standard error as defined in Extended Data Fig. 10. Solid curves: best fits using the numerical simulations. The fits match $f_c$ (yielding $\sigma_{//}$) and the y-axis position of the high-$f$ plateau (yielding $\varepsilon_{//}$). Dashed curves: same simulations for bulk water ($\varepsilon_{bulk} = 80$ and $\sigma_{bulk} = 2\cdot10^{-4}$ S·m$^{-1}$; see Methods). The black lines show the evolution of $f_c$ and $f_r$ with decreasing $h$. (**b**) Dielectric images taken for individual channels with $h \approx 1.5$ and 5 nm (empty and filled with water), as specified in the legends. Upper panels: images for a fixed-$y$ position of the AFM tip; ~ 25 lines were recorded at each $f$. Lower panels: images at 1.1 GHz, ~ 100 lines are shown for clarity because of higher noise. Scale bars, 500 nm.

The measured spectra are essentially determined by the two characteristic frequencies $f_r$ and $f_c$ that in turn reflect $\varepsilon_{//}$ and $\sigma_{//}$ of nanoconfined water, respectively. Accordingly, changes in the frequencies with varying $h$ can be translated into relative changes of the corresponding electrical characteristics. With reference to the inset of Fig. 1a, the experimental spectra in Fig. 2a allow one to conclude immediately that both $\varepsilon_{//}$ and $\sigma_{//}$ increase with decreasing the water thickness. However, to find the scale of these changes and the absolute values of $\varepsilon_{//}$ and $\sigma_{//}$ requires knowledge of the factor $\alpha$ that depends on $h$ and other geometric parameters. To this end, we used full-3D numerical simulations and calculated the detailed electric field distribution across our nanochannels. From this analysis, we obtained anticipated dielectric dispersions for different $h$ (Extended Data Figs. 5 and 6). Then, we determined $\sigma_{//}$ and $\varepsilon_{//}$ by fitting $f_c$ and positions of the high-$f$ plateau (all other parameters were found experimentally; Extended Data Fig. 10). These simulations also took into account that the dielectric response of nanoconfined water is anisotropic, using the $\varepsilon_\perp(h)$ dependence reported previously[22]. Examples of this analysis are presented in Fig. 2a by the solid curves. For comparison, the dashed curves in the same figure represent the behavior expected if the confined water retained the electrical properties of bulk water. In addition, we analyzed the experimental spectra using two analytical approximations where i) the AFM tip was simulated as a point charge and the nanochannel as a laterally infinite multilayer stack (Extended Data Fig. 8) and ii) the measurement geometry was simulated using an equivalent electrical circuit (Extended Data Fig. 9). This provided an approximation for $\alpha$ as a function of the main parameters, $\alpha \cong (h \cdot w/l^*) \cdot [1/(2\pi R) + H/(\varepsilon_{hBN} \cdot w \cdot l^*)]$, where $R$ is the AFM tip radius and $l^*$ is the effective channel length (Methods). Although less accurate, the latter models were found to yield close values for both $\varepsilon_{//}$ and $\sigma_{//}$.

Furthermore, we verified that the extracted values depended little on $\varepsilon_\perp$ (because $\varepsilon_{//}$ is much larger, see Methods, Extended Data Fig. 6f and Extended Data Fig. 7) and were independent of $\sigma_\perp$ (the employed measurement geometry is insensitive to the latter, see Extended Data Fig. 6e). The use of several approaches have shown that our results are robust with respect to modeling and, also, corroborated our qualitative conclusions drawn from the spectral shifts, as discussed above.

We have studied many nanochannel devices with $h$ ranging from ~1 to 60 nm and, using $f$ from a few hundred Hz to GHz, obtained their dielectric spectra similar to those shown in Fig. 2a. Then applying the full numerical analysis for each device, water's $\varepsilon_{//}$ and $\sigma_{//}$ were extracted as a function of $h$. The results are summarized in Fig. 3.

**In-plane dielectric constant**
Inside our channels with $h \geq 5$ nm, water was found to exhibit $\varepsilon_{//} \approx 74 \pm 17$, that is, little difference with respect to bulk water (Fig. 3a). The dielectric constant suddenly changed for quasi-2D water ($h \approx 1$-2 nm) that showed a sharp increase in $\varepsilon_{//}$ by an order of magnitude, reaching $1{,}030 \pm 350$. Such large dielectric constants are typical of ferroelectrics. This behavior is in stark contrast to that of $\varepsilon_\perp$ for the same range of thicknesses[22]. The latter constant exhibits a decrease by a factor of ~40 so that quasi-2D water becomes essentially nonpolarizable in the perpendicular direction[22]. Although the giant $\varepsilon_{//}$ for water under atomic-scale confinement may seem surprising, the result agrees (at least qualitatively) with most simulations that have predicted enhanced $\varepsilon_{//}$ and strong dielectric anisotropy for water in the vicinity of solid interfaces[23,24,27-30]. The phenomenon can be understood by noting that, while the interaction with solid surfaces hinders water dipoles' ability to align along $E$ applied in the perpendicular direction, the dipoles are still relatively free to rotate and align in the in-plane direction.

For large channels, it is reasonable to consider confined water as consisting of an inner layer of bulk water molecules unaffected by the confinement and two near-surface layers with a different interfacial dielectric constant $\varepsilon_{//int}$, as sketched in the inset of Fig. 3a. This simple model (three capacitors in series) has previously proven successful in describing the observed $\varepsilon_\perp(h)$ dependence qualitatively well and suggested the presence of an 'electrically dead' near-surface layer[22]. Its thickness $h_{int}$ was estimated as ~7 Å, in agreement with the thickness of layered water at solid interfaces[35]. However, if the same model is applied to explain the in-plane response [three capacitors in parallel, yielding the effective $\varepsilon_{//}(h) = \varepsilon_{bulk} + 2(\varepsilon_{//int} - \varepsilon_{bulk})h_{int}/h$], the observed $\varepsilon_{//}(h)$ dependence cannot be described even qualitatively (light red dashed curve in Fig. 3a). The sharp rise at $h \approx$ 2-3 nm clearly indicates a transition in the dielectric properties of water if the two confining surfaces get sufficiently close so that the near-surface layers merge and water develops a more distinctly layered structure[4,36]. The large increase in $\varepsilon_{//}$ in this case is consistent with theory[28,29], although the transition was envisaged for $h \gtrsim$ 1 nm. Let us also mention the slight tendency seen in Fig. 3a at $h >$ 4 nm for $\varepsilon_{//}(h)$ to increase with decreasing $h$. This increase would be consistent with many predictions[27-30] and, within the 3-layer model and our experimental accuracy, suggests values of $\varepsilon_{//int}$ up to ~150 for interfacial water. Measurements at higher accuracy are required to corroborate the latter observation.

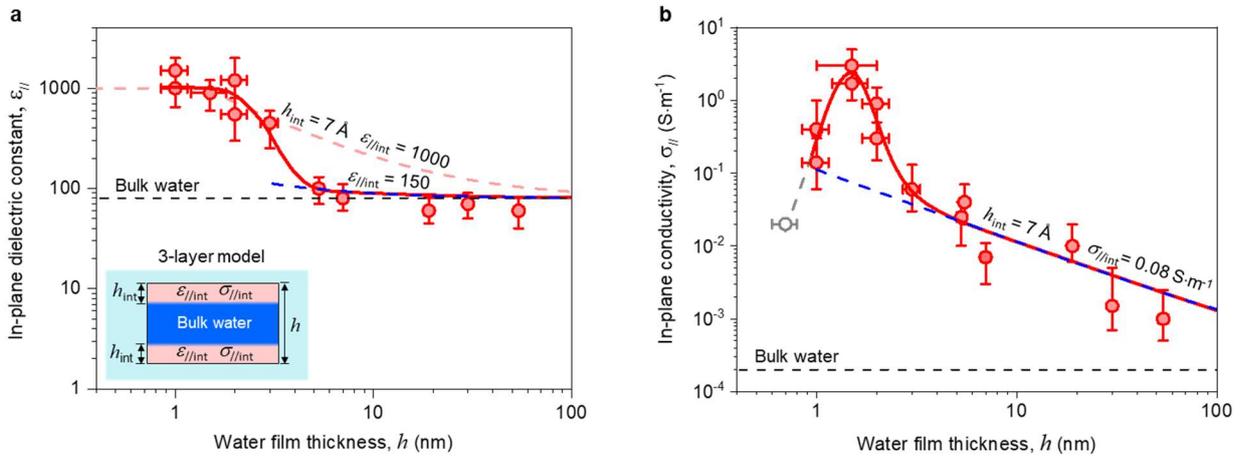

**Fig. 3| In-plane electrical properties of water under molecular-scale confinement. (a)** Experimental data for $\varepsilon_{//}$ are shown by red symbols. Solid curve: guide to the eye. Long-dash curves: 3-layer capacitance model explained in the inset. Shown in light red and blue are the model's fits with $\varepsilon_{//int}$ = 1,000 and 150, respectively, using the interfacial-layer thickness $h_{int}$ found in Ref.[22] Black horizontal line, $\varepsilon_{bulk}$. **(b)** Red symbols and red curve: same as in (a) but for $\sigma$. Grey symbol: experimental value incorporated from Ref.[33] Black line: bulk conductivity of the used water. Blue line: best fit for $h \geq$ 4 nm using the 3-layer model in (a) and assuming different conductivities for bulk and near-surface water. The error bars in both panels are the uncertainty as defined in Extended Data Fig. 10.

**In-plane conductivity**
The extracted conductivities for nanoconfined water are plotted in Fig. 3b. One can see that, for moderate confinement, $\sigma_{//}(h)$ evolves approximately as $1/h$ increasing from $10^{-3}$ - $10^{-4}$ S·m$^{-1}$ for our largest channels up to ~0.1 S·m$^{-1}$ at $h \approx$ 3 nm. This gradual increase is well described by the discussed 3-layer model (inset of Fig. 3a) assuming unaffected conductivity for the inner layer and the presence of two near-surface layers with interfacial conductivity $\sigma_{//int}$, which yields $\sigma_{//}(h) = \sigma_{bulk} + 2(\sigma_{//int} - \sigma_{bulk})h_{int}/h$. Using the same $h_{int}$ for interfacial water as for the dielectric constant, we estimate $\sigma_{//int}$ as ~0.1 S·m$^{-1}$, three orders of magnitude higher than $\sigma_{bulk}$ ($\sigma_{bulk}$ was measured independently using our bulk water with pH $\approx$ 5.7 caused by $CO_2$ absorption). Such an enhancement of confined water's conductivity agrees well with previous reports for water inside nanochannels and is attributed to the surface charge[15,16,19,33,34].

For quasi-2D water, where the water becomes layered across its entire thickness ($h \gtrsim$ 2 nm), we observe a pronounced additional peak in conductivity. The behavior clearly diverges from the surface charge (3-layer) model but correlates well with the anomalous increase in $\varepsilon_{//}$ (cf. Figs. 3a and 3b). At its peak, $\sigma_{//}$ reaches ~3 S·m$^{-1}$, which occurs at $h \approx$ 1.5 nm and corresponds to 4-5 molecular layers[36]. This value is one of the highest conductivities reported at RT, comparable to that of Nafion and approaching the superionic levels reached for molten salts (~ 10 S·m$^{-1}$)[37]. Note that our smallest channels ($h \approx$ 1.0 nm) exhibited a marked (> 10 times) decrease in $\sigma_{//}$ with respect to the peak value. To corroborate the presence of the peak for quasi-2D water, Fig. 3b incorporates a data point available in the literature (grey symbol), which was obtained for water inside 0.7 nm hBN channels using DC measurements[33]. This also shows consistency between different types of experiments.

**Discussion**
Our experiment reveals that water under confinement exhibits two distinct regimes. Under moderate confinement ($h$ down to ~ 2-3 nm), water can be described as effectively containing inner bulk water and interfacial water. The latter is characterized by enhanced conductivity (by at least 3 orders of magnitude for the case of hBN surfaces) and a highly anisotropic dielectric constant with $\varepsilon_\perp \approx$ 2 and $\varepsilon_{//}$ comparable to $\varepsilon_{bulk}$ or slightly larger. For the extreme, atomic-scale confinement that allows only a few molecular layers inside, the quasi-2D water clearly switches into another state. It is different from either bulk or interfacial water and exhibits ferroelectric-like polarizability and superionic-like conductivity (presumably provided by proton transport).

The electrical behavior observed under the moderate ($\gtrsim$ 3 nm) confinement generally agrees with the previous reports and theoretical expectations[15,19,27-30]. As for the atomic-scale confinement, both measurements[4] and molecular dynamic simulations[36] have previously revealed that water has a distinctive layered structure exhibiting more profound density oscillations than those found for near-surface water. The observed anomalies in the electrical properties of few-layer water have not been anticipated in theory, although giant, ferroelectric-like values of $\varepsilon_{//}$ were suggested in Refs.[28,29] and superionic conductivity was predicted for monolayer water at high temperatures[38]. Furthermore, experiments on water flow and ionic transport also reported the emergence of anomalies in the same few-layer regime[33,39].

In the absence of a theory, we note that atomic-scale confinement induces strong hydrogen-bond disorder because of constraints on water molecules' possible orientations and, also, a loss in hydrogen bonding imposed by the two confining surfaces. Qualitatively, this suggests changes in the electrical properties analogous to those found for disordered hydrogen-bonded crystals[40] and in particular in disordered ice. The latter is known to exhibit higher $\varepsilon$ than that of liquid water (and ordered ice) due to easier water molecules' reorientation[14]. Similarly, we speculate that the disruption of the hydrogen-bond network in quasi-2D water enables correlated reorientations of molecular dipoles, allowing their collective polarization[41], and consequently higher, ferroelectric-like $\varepsilon_{//}$[12,28]. Water dipoles' reorientation is essential in the Grotthuss mechanism that explains fast proton transport in water. Accordingly, correlated dipolar reorientations in quasi-2D water should also facilitate faster proton transfer, resulting in higher $\sigma_{//}$, on top of the increase in near-surface proton concentration due to surface charges.

**Conclusion**
We have measured the in-plane electrical properties of 2D nanoconfined water and found that they radically change as the water's layered structure becomes more pronounced under extreme, molecular-scale confinement. We expect this behavior to be general for strongly confined water and not limited to the case of hBN channels studied here. Our results are fundamental to interpreting numerous physical and biochemical processes occurring at the molecular scale. In particular, they provide insights into the electric double layer under strong confinement, which is essential for developing energy storage and generation technologies. The reported approach to measure conductivities and dielectric constants can be applied to substances other than water and has the notable advantage that it probes local properties without averaging over a distribution of different confinement strengths as inevitable in macroscale measurements.

**METHODS**

**Device fabrication.** We fabricated our devices using procedures similar to those described in Ref.[22] A free-standing silicon nitride (SiN) membrane was used as a substrate for our vdW assembly that involved four atomically-flat crystals (Extended Data Fig. 1). We started with etching a rectangular aperture of ~3×20 $\mu m^2$ in the SiN membrane. This aperture was required to serve later as a water inlet from a reservoir placed at the back of the wafer. Using the dry transfer method, we transferred a graphite crystal (thickness ~ 10 nm) to seal the aperture. The graphite later served as the bottom electrode. It was etched through from the backside using reactive ion etching, which projected the aperture into graphite. Then a relatively thick hBN crystal ($H \approx$ 50 - 200 nm; to serve as the bottom layer) was transferred on top of the graphite and again dry-etched through from the back. Next, we selected a second hBN crystal (thickness $h \approx$ 1 - 60 nm), referred to as spacer, and patterned it into parallel stripes (spaced by ~ 200 nm) using e-beam lithography and reactive ion etching. We transferred the spacer layer on top of the bottom hBN using wet transfer procedures and aligning the stripes perpendicular to the rectangular aperture. Finally, the third hBN crystal was transferred on top of the spacer layer using the dry transfer method. This sealed the nanochannels as well as the aperture. The thickness of the top hBN ($H_{top} \approx$ 20 - 80 nm) was carefully chosen in order to bring the AFM tip as close as possible to the water but also to ensure that the top layer exhibits some sagging into empty channels without blocking them completely[22] (Extended Data Figs. 2a and 2d). After each transfer, we annealed the assembly in Ar/$H_2$ at 300°C for 3 hours and then at 400°C for 5 hours to remove polymer residues and other contamination. As the final step, we made an electrical contact with the graphite using photolithography or, alternatively, using silver paint to minimize the number of cleanroom processes. Because the top layer became flattened (no longer sagged)[22] after liquid water filled the channels (see Extended Data Figs. 2d-g), this allowed us to ensure that, irrespective of water's dielectric response, the probed nanochannels were fully filled with water (by comparing AFM topography images before and after filling; Extended Data Figs. 2e and 2f). Note that, given the large dielectric response of water found in this work for all channel's thicknesses, we could also verify the water filling by taking AFM images in the intermittent-contact attractive mode using low-$f$ applied voltages (a few kHz down to DC). In this case, the acquired contrast over the channels reversed from negative to positive upon filling the water (Extended Data Figs. 2b and 2c) due to the onset of a strong electrostatic force associated with water's conductivity.

As another improvement with respect to our previous studies, before filling with water, we normally exposed the bottom side of the devices to low-power Ar/$O_2$ plasma (8 W, 16 sccm Ar and 8 sccm $O_2$ flow) for 1 sec. This made SiN more hydrophilic and also cleared channels' entrances from possible contamination. This procedure has proven beneficial for getting water inside. We also found that our devices tended to delaminate if contamination remained trapped between vdW layers. To prevent this from happening, great care was taken to check for cleanliness of the devices at each fabrication step (using optical and AFM imaging). For example, if dark-field optical images showed bubbles trapped between layers, such devices were discarded. Representative optical images of the studied devices with clean interfaces are provided in Extended Data Figs. 1a-c.

**Local broadband dielectric imaging and spectroscopy.** All SDM[10,42] measurements were carried out using a commercial AFM (Nanotec Electronica with WSxM software[43]) operated at RT in dry atmosphere. SDM was implemented in the amplitude-modulated electrostatic-force detection mode[44,45], adapting the approach described in Ref.[10] Briefly, we applied an AC voltage between the AFM tip and the bottom electrode. By detecting mechanical oscillations of the cantilever, we measured the electrostatic force and, therefore, the first derivative of the tip-sample capacitance, $dC/dz$. Its value depends on both dielectric and conductive properties of probed samples. SDM has previously been used for local measurements of both surface and sub-surface dielectric properties of various materials in different frequency regimes, from quasi-static to GHz[46-49]. The same approach has also been widely used to study near-surface electron transport in solid samples[50-53], showing the technique's sensitivity to local conductivity. The force-sensing approach was preferable in our case to the other current- and microwave-sensing AFM techniques that can also probe local impedance[54,55] because the latter are generally less sensitive and more complex to implement.

To carry out dielectric spectroscopy over the required very wide bandwidth (100 Hz – 1 GHz), we combined two previously reported force-sensing detection methods. Both were implemented here using conductive diamond-coated AFM probes (CDT-CONTR or CDT-FMR from Nanosensors™) with spring constant $k$ of few Nm$^{-1}$ and resonance frequency in the range of 20-60 kHz. At low $f$ up to the cantilever resonance frequency, we used the 2$\omega$-detection approach, as described in Ref.[22] for measurements of water's $\varepsilon_\perp$, where $\omega = 2\pi f$ is the angular frequency. Briefly, we applied an AC voltage with amplitude $v_{AC}$ = 4 V and measured the amplitude $D_{2\omega}(z)$ of the resulting mechanical oscillations of the cantilever at double the applied frequency, using an external lock-in amplifier (Zurich Instruments HF2LI). The AFM tip-sample capacitance gradient was calculated as $|dC/dz| = D_{2\omega}(z) \cdot 4k/v_{AC}^2$. For $f$ higher than the cantilever resonance frequency, we used the heterodyne detection technique demonstrated in Ref.[56] To this end, we applied a high-$f$ (carrier) signal modulated by low frequency ($f_{mod}$ = 1 kHz, amplitude $v_{AC}$ = 0.5 V) using an external RF/GHz signal generator (Rohde & Schwarz SMA100B). We detected the cantilever mechanical oscillations at $f_{mod}$ using our external lock-in amplifier and, from those measurements, obtained the capacitance gradient at the carrier $f$ as $|dC/dz| = D_{mod}(z) \cdot 8k/(g \cdot v_{AC}^2)$, where $g$ is the $f$-dependent gain of the external circuit, $D_{mod}(z)$ is the amplitude of the cantilever mechanical oscillations at $f_{mod}$. This approach allowed retrieval of $|dC/dz|$ variations at $f$ higher than the cantilever resonance frequency and up to GHz frequencies[52]. Note that such measurements yield only the amplitude of $dC/dz$ and not its phase, so that the obtained spectra reflect the modulus of the complex dielectric constant (see "Analysis of local dielectric spectra").

To minimize systematic errors, for each device and each AFM tip, we carried out careful calibration of the electrical gain $g$ at each measurement frequency. This is essential especially at high frequencies (> 1 MHz), as both applied and measured signals are subject to large $f$-dependent variations because the cable impedance is not matched to the local sample impedance and the long-range impedance between the AFM tip and the sample. The gain factor was calibrated using a similar procedure to that introduced in Ref.[48], which relies on acquiring $|dC/dz|$ curves as a function of the tip-surface distance $z$ over a region of the device that exhibits $f$-independent impedance. We then determined $g$ by comparing $|dC/dz|(z)$ curves at specific frequencies against a low-$f$ (reference) curve, where no gain or loss is expected. To minimize possible changes in the gain in different regions of the device, we took the $|dC/dz|(z)$ curves at various $f$ over the center of the hBN spacer close to the studied water-filled channel (Extended Data Fig. 3a). We then scaled them to match the curve taken at 2 kHz using the gain $g$ and the offset as two fitting parameters (Extended Data Fig. 3c). The latter is always present in $dC/dz$ curves, irrespective of frequency, and is typically subtracted before analysis, because the offset does not depend on local electrical properties (see, e.g., Refs.[10,57]). Indeed, the measured offsets were found to change substantially at high $f$ (10 MHz – 1GHz), but they were the same above water channels and hBN spacers. This confirmed that, for all the frequencies, the offsets originated from long-range impedance contributions. Importantly, while $g$ is critical for accurately evaluating the electrical properties of water from the measured signals, offsets play little role. This is because we analyzed relative changes in $|dC/dz|$ along the device surface, that is, water channels vs hBN spacers (see "Dielectric images and dielectric spectra acquisition"). We verified our calibration procedure using reference samples (hBN on doped Si) and obtained good agreement with the expected $f$-independent dielectric spectra up to GHz frequencies (Extended Data Fig. 3e,f). Furthermore, large channels filled with water effectively served as another reference, and the obtained calibrated spectra exhibited the expected behavior (flat response at high $f$) and the dielectric constant of bulk water. We note that if the separation between channels is too small, calibration curves obtained above hBN spacers could, in principle, contain an electrostatic contribution coming from water itself. This would in turn lead to an underestimate of the dielectric constant of water inside the channels. To avoid such crosstalk, we have chosen the distance between the channels to be ~800 nm, large enough. Using numerical simulations, we also checked that, within our experimental accuracy (~1 zF·nm$^{-1}$), the calibration curves obtained in our setup were not influenced by electrostatics coming from water inside the channels (Extended Fig. 7f).

**Dielectric images and dielectric spectra acquisition.** The dielectric images in Fig. 1 and Extended Data Fig. 4 were taken at the constant height $z_{scan}$ (typically ~ 15-25 nm) from the top layer, as in Ref.[22] The images as a function of $f$ reported in Fig. 2b were taken at both constant $z_{scan}$ and constant $y$ position ($y$-axis is along the nanochannels length). We typically recorded 25 lines at each $f$. Standard AFM image processing consisting of flattening and gaussian filtering was applied. The dielectric spectra in Fig. 2a were obtained for constant $z_{scan}$ and constant $y$, but after acquiring the entire 3D data set $|dC/dz|(x,z,f)$ such as shown in Extended Data Fig. 3d. This set is composed of many $|dC/dz|(x,z)$ images (see Extended Data Fig. 3a), which were obtained at different $f$ by scanning the AFM tip across the channels (along the $x$-axis) at constant $y$ and approaching the surface in steps. This procedure allowed us to minimize errors in $z_{scan}$ due to vertical drifts. When taking $|dC/dz|(x,z)$ images, the AFM tip was approached from large distances towards the surface (down to ~15 nm). At each step, we acquired dozens of lines, which were then averaged to obtain the corresponding profile (Extended Data Fig. 3b). This notably improved the signal-to-noise ratio. The vertical drift of the AFM tip was compensated by comparing the $|dC/dz|(z)$ curve extracted from the 3D data set with a rapidly measured $|dC/dz|(z)$ curve taken above the hBN spacer region immediately before/after the imaging. By taking small steps near to the surface, we could then reconstruct the dielectric spectra (such as in Fig. 2a of the main text) by using the value measured in the middle of nanochannels at the same $z_{scan}$ (± 1 nm) for each $f$. In addition to the scanning height $z_{scan}$, the spectral values depended on geometry/size of our devices and of AFM tip's parameters (Extended Data Fig. 10). All the necessary geometric parameters of the studied devices were measured by taking their topography images, whereas the AFM tip parameters (its radius $R$ and half-angle $\theta$) were determined by fitting $|dC/dz|(z)$ curves taken directly above the graphite bottom layer, as in Ref.[22], following the procedures described in Refs.[57,58]. For example, the tip radii were found in the range 50-200 nm, in good agreement with the values specified for the commercial diamond-coated tips.

**Numerical modeling and data analysis.** The observed dielectric spectra were fitted to 3D finite-element numerical calculations implemented using COMSOL Multiphysics 5.4a (AC/DC electrostatic module). These calculations were based on the electrostatic model previously used in Ref.[22], adapted here to simulate the frequency-dependent force acting on the tip as a function of $\varepsilon$ and $\sigma$ of water if an AC electric field was applied. A schematic of the model is shown in Extended Data Fig. 5a. Following Ref.[22], the AFM tip was modeled as a truncated cone with the half-angle $\theta$ terminated with a tangent hemispherical apex of radius $R$. The values of $\theta$ and $R$ used in our simulations were measured for each AFM tip, as discussed above. The AFM cone height $H_{cone}$ was limited to 6 µm and the cantilever was omitted to reduce the computational time. We checked that these approximations had no impact on the simulated results. Nanochannels were simulated as 3 rectangular parallelepipeds of length $l$ = 3 µm (Extended Data Fig. 5a). Their height $h$, width $w$ and spacing $w_s$ were determined from topography imaging for each device. The water channels' electrical response was modeled with the anisotropic, complex dielectric constant

$$\varepsilon^*_{\perp,\parallel}(\omega) = \varepsilon_0 \varepsilon_{\perp,\parallel} - i \frac{\sigma_{\perp,\parallel}}{\omega}, \quad (1)$$

where $\omega$ is the angular frequency, $\varepsilon_0$ is the dielectric permittivity of vacuum, $\varepsilon_\perp$ and $\varepsilon_{\parallel}$ are the dielectric constants perpendicular and parallel to the channel, respectively, and $\sigma_\perp$ and $\sigma_{\parallel}$ are the perpendicular and parallel conductivities, respectively. Note that in Eq. (1) $\varepsilon_{\perp,\parallel}$ can contain ionic contributions due to ion-water and ion-ion correlations in addition to water's purely dielectric response[59,60].

The dielectric media surrounding the nanochannels (Extended Data Fig. 5a) was modeled as a rectangular cuboid extending above and below the water channels (length $l$ = 3 µm, width $W$ = 3 µm, and height $H_{top} + h + H$), which reflected the corresponding dimensions of the top, bottom and spacer hBN layers. The anisotropic dielectric constant of hBN was set to its bulk values ($\varepsilon_{\perp hBN} \approx 3$, $\varepsilon_{\parallel hBN} \approx 5$). For each device, we numerically solved the Poisson's equation in the frequency domain and calculated the force acting on the AFM probe and, from that, $|dC/dz|$ by integrating the built-in Maxwell stress tensor on the probe's surface using the same simulations' box size and boundary conditions as in Ref.[22]. Note that, for brevity, in all the figures the notion $|dC/dz|$ refers to the relative dielectric contrast, that is, the response relatively to the hBN spacer so that $|dC/dz| \equiv dC(z_{scan},\varepsilon_{\perp,\parallel},\sigma_{\perp,\parallel})/dz - dC(z_{scan}, \varepsilon_{\perp,\parallel hBN},0)/dz$. Accordingly, we computed $|dC/dz|$ over the center of the water channel as a function of $f$ and subtracted the corresponding values for the case of the tip placed over the center of the hBN spacer. Examples of the calculated dielectric spectra for representative geometrical parameters of our devices and water's $\varepsilon$ and $\sigma$ are plotted in Extended Data Fig. 6. The plots closely reproduce the experimental behavior (for details, see "Analysis of local dielectric spectra"). For each device, we extracted water's $\varepsilon_{\parallel}$ and $\sigma_{\parallel}$ by fitting the calculated dispersions to the experimental one. Note that $\varepsilon_{\parallel}$ and $\sigma_{\parallel}$ were the only unknowns in our analysis, as all the other parameters needed for the simulations were determined experimentally. $\varepsilon_\perp$ was set to the values measured in Ref.[22] and $\sigma_\perp$ was set to the measured value for our bulk water ($\sigma_{bulk}$ = 2 x 10$^{-4}$ S·m$^{-1}$). We also found that the extracted results for $\varepsilon_{\parallel}$ and $\sigma_{\parallel}$ depended little on exact values of $\varepsilon_\perp$ or $\sigma_\perp$, showing that our experimental geometry was rather insensitive to water's out-of-plane characteristics, and the influence was negligible for our smallest channels (Extended Data Figs. 6e,f and 8e,f). This can be understood as being due to the fact that the nanochannel's impedance in the out-of-plane direction is much smaller than the impedance in series due to the hBN layers and AFM tip-surface capacitances, whereas the nanochannel's impedance in the in-plane direction is much larger (see "Electrical modeling" below).

It is important to note that the thickness $H$ of the hBN bottom layer determines whether the measurements are mostly sensitive to $\varepsilon_{\parallel}$ or $\varepsilon_\perp$. Extended Data Fig. 7 illustrates this for the case of channel thickness $h$ = 5 nm at high $f$, beyond the conductivity relaxation regime (see "Analysis of local dielectric spectra") and for a large AFM tip radius (100 nm) as used in our experiments. When the probed channel is in immediate proximity to a metallic surface (Extended Data Fig. 7a), as in Ref.[22], the $|dC/dz|$ signal is independent of $\varepsilon_{\parallel}$ and, consequently, the extracted $\varepsilon$ represents the out-of-plane component, $\varepsilon_\perp$. Furthermore, measurements are sensitive only to relatively small values of $\varepsilon_\perp$ (up to ~20 for $h$ = 5 nm), as the signal saturates with further increases in $\varepsilon_\perp$ (for larger channel thicknesses, the sensitivity extends to increasingly larger values of $\varepsilon_\perp$, as shown in Ref.[22]). Conversely, when the channel is on a thick hBN layer ($H$ = 200 nm; Extended Data Fig. 7b), the $|dC/dz|$ signal significantly increases for $\varepsilon_{\parallel}$ > 20, becoming dominated by the in-plane component. Therefore, for relatively large values of $\varepsilon_{\parallel}$ ranging from ~80 to ~1,000, as measured in this work, the signal at such $H$ shows little dependence on $\varepsilon_\perp$. Extended Data Fig. 7c shows how the signal evolves as a function of $H$. In these simulations, we increased the cone height and cantilever length of the AFM tip to their nominal values ($H_{cone}$ = 12 µm and $L_{cantilever}$ = 20 µm). This allowed us to include electrostatic contributions from longer-range components of the AFM probe[61], which we found to be relatively small but still non-negligible for very thick bottom hBN ($H$ > 500 nm). Results show that the $|dC/dz|$ signal and its sensitivity to $\varepsilon_{\parallel}$ peaks between 50 and 500 nm (Extended Data Fig. 7c). For thinner hBN layers, the $\varepsilon_{\parallel}$ contribution is relatively small or negligible, and the signal is dominated by $\varepsilon_\perp$. For hBN thicker than 500 nm, the $|dC/dz|$ signal decreases becoming less sensitive to the probed local electrical properties, consistent with previous results[57]. This is expected because the AFM tip is moved far away from the bottom electrode. In this work we used bottom hBN layers of thickness up to 200 nm. This limitation was dictated by constraints in fabrication of our devices, because thicker hBN crystals were stiffer and provided poor adhesion, leading to delamination upon filling the channels with water.

**Analytical modeling.** Because of the nonuniform distribution of electric field across nanochannels and the complex geometry of AFM probes, the electrostatic problem cannot be solved exactly using analytical models. Hence, we used the 3D numerical simulations described above. Nonetheless, it is rather informative to also provide an analytical approximation that would substantiate our numerical results and could be used for semi-quantitative estimates. To this end, we employed the point-charge model where the AFM tip was replaced by a point charge $Q$ positioned at distance $z = z_{scan} + R$ from the top hBN surface (Extended Data Fig. 8a). The devices were modeled as a stack of different slabs, infinite in the in-plane direction. The slabs corresponding to the hBN layers were positioned at $0 < z < - H_{top}$ and $- (H_{top} + h) < z < - (H_{top} + h + H)$ and modeled as insulators with zero conductivity. The water layer at $-H_{top} < z < - (H_{top} + h)$ was described by anisotropic dielectric constants $\varepsilon_{\parallel,\perp}$ and conductivities $\sigma_{\parallel,\perp}$. The boundary condition of zero voltage was imposed at $z = - (H_{top} + h + H)$ to model the highly conducting ground electrode. By leveraging the in-plane translational invariance, the Laplace equation for the Fourier transform electrostatic potential in the in-plane direction $\phi_q(z)$ becomes

$$q^2 \varepsilon_{\parallel}(z)\phi_q(z) - \partial_z[\varepsilon_\perp(z)\partial_z\phi_q(z)] = Q_q\delta(z - z_{scan} - R) \quad (2)$$

where $\mathbf{q} = (q_x, q_y)$ is the wave vector in the in-plane direction, and the position-dependent dielectric constants are denoted as $\varepsilon_{\parallel}(z)$ and $\varepsilon_\perp(z)$, respectively. The dielectric constant of the hBN slabs was, for simplicity, assumed isotropic with $\varepsilon_{\parallel}(z) = \varepsilon_\perp(z) = \varepsilon_{hBN} = 4$. The Laplace equation was solved within each slab using exponential functions. The solutions were matched at each interface by imposing the following boundary conditions: continuity of $\phi_q(z)$ and

$$\varepsilon_\perp(z+\eta)\partial_z\phi_q(z+\eta) - \varepsilon_\perp(z-\eta)\partial_z\phi_q(z-\eta) = 0 \quad (3)$$

where $\eta$ is an infinitesimal constant. A similar relation holds at the point-charge location, $z_{scan} + R$, but the right-hand side in (3) becomes equal to $Q$. The potential in real space $\phi(r,z)$ was then obtained by performing the Fourier transform of $\phi_q(z)$ in the in-plane direction. The capacitance was estimated as $C = Q/\phi(0, z_{scan})$, and from the latter the capacitance gradient $|dC/dz|$ was calculated. Examples of the calculated dielectric spectra for representative parameters of our devices and water's $\varepsilon$ and $\sigma$ are given in Extended Data Fig. 8, where again for brevity $|dC/dz|$ indicates its value relatively to the case of the heterostructure fully made of hBN, $|dC/dz| = dC(z_{scan}, \varepsilon_{\perp,//}, \sigma_{\perp,//})/dz - dC(z_{scan}, \varepsilon_{hBN}, 0)/dz$. The calculated spectra show that the model captures all the main features of the observed behavior as a function of various parameters and agrees well with our numerical simulations in Extended Data Fig. 6. Note however that, because the analytical model assumes an infinite water layer and does not account for a finite $w$, the transition between low and high $f$ plateaus becomes somewhat less pronounced (Extended Data Fig. 8d). The analytical results agree with the numerical simulations only for $w$ much larger than the tip radius $R$ (Extended Data Fig. 6d). Accordingly, if only the analytical model were used to fit the reported experimental data, this would result in systematic underestimation of both $\varepsilon_{//}$ and $\sigma_{//}$. In particular, the values extracted for quasi-2D water layers ($h \gtrsim 2$ nm) would be underestimated by a factor of about 5, although all the trends with changing $h$ would remain correct.

**Analysis of local dielectric spectra**. The observed dispersion came from the DC conductivity of water, not from its dipolar orientational (Debye) relaxation[62]. The latter occurs only at $f \approx 10$ GHz in bulk water[63], well above the frequencies considered in our report. The reason for a DC conductivity contribution into our spectra can be understood by noticing that we measured the modulus of $dC/dz$, that is, the modulus of the complex dielectric constant $|\varepsilon^*_{\perp,\|}(\omega)|$. As a result, the dielectric response is determined by both $\varepsilon$ and $\sigma$ and, depending on frequency, either the first or second term in Eq. (1) becomes dominant. At sufficiently low $f$, the conductivity term always dominates $|\varepsilon^*_{\perp,\|}(\omega)|$, similar to the case of macroscale measurements using standard broadband dielectric spectroscopy[64]. To this end, it is useful to recall that at low frequencies the spectrum of deionized water at macroscale is known[63] to be dominated by $\sigma_{bulk}$. Therefore, the spectrum is effectively divided into two regions separated by the conductivity relaxation frequency, $f_{r,bulk}$, at which the behavior changes. In the conduction-dominated region ($f < f_{r,bulk}$), the dielectric response decreases with increasing $f$, while for $f > f_{r,bulk}$ it is constant and depends only on $\varepsilon_{bulk}$. The value of $f_{r,bulk}$ is given by[64]

$$f_{r,bulk} = \frac{\sigma_{bulk}}{2\pi\varepsilon_0\varepsilon_{bulk}} \quad (4)$$

which follows from Eq. (1) if we use $|\varepsilon_{\perp,\|}| = \varepsilon_{bulk}$ and $|\sigma_{\perp,\|}| = \sigma_{bulk}$. Note that this frequency is analogous to the cutoff frequency for systems that can be modeled by a simple RC circuit (see "Electrical modeling"). For the bulk water used in our experiments ($\varepsilon_{bulk} \approx 80$, $\sigma_{bulk} \approx 2 \cdot 10^{-4}$ S·m$^{-1}$), Eq. (4) yields $\sim 45$ kHz. This value agrees well with $f_{r,bulk}$ obtained from both numerical and analytical modeling discussed above for the specific experimental geometry. Indeed, the analyses shown in Extended Data Figs. 6b and 8b yielded the purely dielectric behavior characterized by high-$f$ plateaus starting at $f \gtrsim 45$ kHz for all our nanochannels, irrespective of their height $h$. The additional plateau found in our simulations at low $f$ reflects the presence of non-lossy dielectrics (air gap between the AFM tip and the device; the top and bottom hBN layers). These dielectrics can be represented as additional capacitances in series to water's contribution ("Electrical modeling"). Accordingly, for $f$ below the cut-off frequency $f_{c,bulk}$, the modeled response becomes purely dielectric, reflecting the series capacitances. For $f_{c,bulk} < f < f_{r,bulk}$, the response is dominated by water's $\sigma$ whereas above the relaxation frequency $f_{r,bulk}$ it is again purely dielectric but now dominated by water's $\varepsilon$. For other relevant values of water's $\sigma$ and $\varepsilon$, the spectra exhibit similar behavior (Extended Data Figs. 6e,f and 8e,f): there is a low-$f$ plateau up to the cut-off frequency $f_c$, above which the response decreases with increasing $f$, and the high-$f$ plateau develops above the conductivity relaxation frequency $f_r$.

The onset of the low-$f$ plateau with decreasing $f$ is determined by the water's in-plane conductivity $\sigma_{//}$ whereas no information can be inferred about $\sigma_\perp$ from our experimental data because the measurement geometry is insensitive to the latter conductivity. This can be seen in Extended Data Figs. 6e and 8e where the low-$f$ plateau shifts in frequency with varying $\sigma_{//}$, but not $\sigma_\perp$, and completely disappears if $\sigma_{//} = 0$ and $\sigma_\perp \neq 0$. Qualitatively, the stronger dielectric response at low $f$ reflects the fact that the electric potential drops mostly between the AFM tip and the conductive water layer. We used our numerical simulations to illustrate this effect (Extended Data Fig. 5). At low $f$ and $\sigma_{//} \neq 0$, the electric potential drops almost entirely across the air and the top hBN layer with little voltage drop left below the water layer (Extended Data Fig. 5e). Also, the potential distribution extends along the length of nanochannel for a few micrometers (Extended Data Fig. 5c). In contrast, at high $f > f_r$, the potential drop occurs across the entire thickness of our devices, quite similar to the case where the AFM tip is placed above hBN spacers (Extended Data Fig. 5e). The potential drop at high $f$ also extends nearly equally in all the lateral directions around the tip apex (Extended Data Fig. 5d), similar to the case of Extended Data Fig. 5b for non-conducting water layer ($\sigma_{//} = 0$). The effective screening by water's conductivity along the channel length explains the observed large low-$f$ response. Note that the absolute value of the low-$f$ response is independent of $\sigma_{//}$ but depends on geometric parameters, in particular the channel width $w$, the tip radius $R$ (Extended Data Fig. 6c) and the bottom-layer thickness' $H$ (Extended Data Fig. 7c and Extended Data Fig. 8c). This makes simulations essential for accurate evaluation of the magnitude of changes on the dispersion curves.

While the value of $\sigma_{//}$ does not affect the low-$f$ response, it controls the cut-off frequency $f_c$, which shifts to higher $f$ proportionally to $\sigma_{//}$ (Extended Data Figs. 6e and 8e) but independently of $\varepsilon_{//}$ (Extended Data Figs. 6f and 8f). This behavior can be described by

$$f_c = \alpha \frac{\sigma_{//}}{2\pi\varepsilon_0}, \quad (5)$$

where $\alpha$ is the geometrical parameter that can be estimated analytically ("Electrical modeling"). For more accurate results, $\alpha$ can also be obtained using numerical simulations (Extended Data Fig. 10). They yielded $\alpha \approx 2.8 \times 10^{-3}$, $6.2 \times 10^{-4}$ and $9.2 \times 10^{-5}$ for our three representative devices discussed in the main text, which had $h \approx 30$, 5 and 1.5 nm, respectively. These values of $\alpha$ suggest that $f_c$ should shift relatively little (from $\sim 10$ kHz to $\sim 500$ Hz) if water filling the channels exhibited the bulk properties (Extended Data Figs. 6b,c and 8b,c). This shift is much smaller than that observed experimentally and, in fact, occurs in the opposite direction, towards lower $f$ for smaller $h$, in stark contrast to the experimental behavior (Fig. 2a of the main text). This reiterates the fact that the observed increase in $f_c$ with decreasing $h$ cannot be associated with changes in geometry but comes from a huge increase in water's $\sigma_{//}$ for stronger confinement.

At high $f$, beyond the conductivity relaxation regime, water behaves as a purely dielectric media and, accordingly, the relative height of the high-$f$ plateau is no longer dependent of $\sigma_{//}$ but depends on water's $\varepsilon$, the channel's geometry (in particular its height $h$), and the AFM tip radius (Extended Data Fig. 6c). The plateau's height allows us to extract $\varepsilon_{//}$ using numerical simulations. Importantly, for small channels and large in-plane dielectric constant ($\varepsilon_{//} \geq 80$), the contribution of $\varepsilon_\perp$ becomes negligible (Extended Data Figs. 6f and 8f). Furthermore, we find that $f_r$ shifts to higher frequencies proportionally to $\sigma_{//}$ and inversely proportionally to $\varepsilon_{//}$, and is given by

$$f_r = \frac{\sigma_{//}}{2\pi\varepsilon_0\varepsilon_{//}}, \quad (6)$$

The equation is independent of the device geometry and presents an equivalent of Eq. (4) valid at macroscale. Therefore, once $\sigma_{//}$ is known, $\varepsilon_{//}$ can be directly obtained from $f_r$ using Eq. (6), provided $f_r$ is well separated from $f_c$ as in the case of our small channels. In addition, if $\alpha$ is also known, both $\sigma_{//}$ and $\varepsilon_{//}$ could readily be estimated from the two characteristic frequencies $f_c$ and $f_r$ without the need of simulations, simply using the equations

$$\sigma_{//} = \frac{2\pi\varepsilon_0 f_c}{\alpha}, \quad (7)$$

$$\varepsilon_{//} = \frac{f_c}{\alpha f_r}, \quad (8)$$

which follow directly from Eqs. (5-6). Note that the above considerations and Eqs. (5-8) can also be helpful for the analysis of dielectric spectra obtained using other scanning probe approaches[45], including those that probe higher derivatives of $|dC/dz|$ and scanning impedance/microwave microscopy that directly probes the local impedance.

**Electrical modeling.** It is instructive to use an equivalent impedance circuit to describe the observed dielectric dispersion. However, because of long-range contributions from various AFM cantilever components and complex geometry of our nanochannel devices that result in a nonuniform distribution of the electric field, an equivalent circuit should be so complicated that it is unrealistic to describe our spectra quantitatively. Below, we provide a simplified model that aims to elucidate the physics behind and support our numerical results (Extended Data Fig. 9). To this end, the AFM tip-nanochannel interaction can be modeled by the capacitance, $C_{tip}$, that for simplicity accounts for both tip-air and top-hBN-layer capacitances and, to a first approximation, can be calculated as $C_{tip} = 2\pi\varepsilon_0 R ln[1+R(1-sin\theta)/(z_{scan}+H_{top}/\varepsilon_{//hBN})]$ using the formula described in Ref.[58] We neglect the stray capacitances associated with the AFM cantilever and consider only the tip apex capacitance that is expected to provide the dominant contribution. As for the water-filled nanochannel, we consider it as a distributed RC network shown in Extended Data Fig. 9a. It consists of two elementary RC circuits describing in-plane and out-of-plane impedances $Z_{//}(\omega) = R_{//}/(1+i\omega R_{//}C_{//})$ and $Z_{\perp}(\omega) = R_{\perp}/(1+i\omega R_{\perp}C_{\perp})$, respectively. $R_{\perp,//}$ and $C_{\perp,//}$ can be estimated as $C_{//} = \varepsilon_0\varepsilon_{//}wh/\Delta l$, $R_{//} = \Delta l/(\sigma_{//}wh)$ for the in-plane direction, and $C_{\perp} = \varepsilon_0\varepsilon_{\perp}w\Delta l/h$ and $R_{\perp} = h/(\sigma_{\perp}w\Delta l)$ for the out-of-plane one, where $\Delta l$ is the length of the circuit element along the channel. We also considered another capacitance $C_b$ in series to $Z_{\perp}$, to model the effect of the bottom hBN layer between the water channel and the ground. Plugging in the experimental values relevant to our devices, one can readily find that $Z_{\perp}(\omega) << 1/(\omega C_b)$ for all frequencies, meaning that the contribution of $Z_{\perp}(\omega)$ is negligible in our experiments, in agreement with the above numerical and analytical calculations. The distributed network can then be simplified further and described by the electrical circuit shown in Extended Data Fig. 9b, where the water impedance is modeled by a single RC unit in the in-plane direction, that is, $Z_{ch}(\omega) = R_{//}/(1+i\omega R_{//}C_{//})$, where $C_{//} = \varepsilon_0\varepsilon_{//}wh/l^*$, $R_{//} = l^*/(\sigma_{//}wh)$ and $l^*$ is the effective length of the nanochannel contributing to electrostatic interactions with the AFM tip. With reference to Extended Data Fig. 5c, $l^*$ notably exceeds the tip diameter and, without loss of generality, can be assumed to be of the order of a few μm. The total equivalent impedance between the AFM tip and ground is then given by

$$Z(\omega) = \frac{1+i\omega R_{//}(C_{//}+C_{geom})}{i\omega C_{geom}(1+i\omega R_{//}C_{//})}, \quad (9)$$

where $C_{geom} = C_{tip}C_b/(C_{tip}+C_b)$ is the capacitance that depends on geometric parameters but not on water's electrical properties. Extended Data Fig. 9c shows $|Z|$ as a function of $f$ for the three representative devices. The effective capacitance of the modeled circuit is given by $1/\omega Z(\omega)$ and displays the same qualitative dependence on $f$, $\sigma$ and $\varepsilon$ of the capacitance gradient, $|dC/dz|$. Extended Data Fig. 9d plots $|1/\omega Z(\omega)|$ that indeed exhibits both low- and high-$f$ plateaus characterized by frequencies, $f_c$ and $f_r$, in good agreement with the experiment and numerical simulations. Despite its simplicity, the electrical model also reproduces well the changes in the high-$f$ plateau with varying water's $\varepsilon_{//}$ (Extended Data Fig. 9f) and changes in $f_c$ with varying $\sigma_{//}$ (Extended Data Fig. 9e).

Using this model, we can also corroborate the expressions for $f_c$ and $f_r$ given by Eqs. (5-6). Indeed, the pole of Eq. (9) yields the relaxation frequency as $f_r = 1/(2\pi R_{//}C_{//})$ and, plugging in $R_{//}$ and $C_{//}$ in terms of $\sigma_{//}$ and $\varepsilon_{//}$, results in Eq. (6). The zero of Eq. (9) yields the cut-off frequency so that $f_c \cong 1/(2\pi R_{//}C_{geom})$ where we take into account that $C_{geom}$ is larger than $C_{//}$. Accordingly, $f_c$ is proportional to $\sigma_{//} \propto 1/R_{//}$ and depends on the measurement geometry (through $C_{tip}$ and $C_b$) but is independent of $\varepsilon_{//}$, in agreement with our numerical simulations. Expressing $C_{tip}$, $C_b$ and $R_{//}$ in terms of geometric and electrical parameters as defined above and taking $C_b = \varepsilon_0\varepsilon_{hBN}wl^*/H$, we obtain Eq.(5) where, in the case of our geometry, the geometric factor $\alpha$ can be approximated to $\alpha \cong (h \cdot w/l^*) \cdot [1/(2\pi R) + H/(\varepsilon_{hBN}wl^*)]$. Using the specific parameters for our representative devices ($h \approx 30$, 5 and 1.5 nm) and the effective channel length $l^* = 3$ μm, this yields $\alpha$ of about $\sim 4 \times 10^{-3}$, $1 \times 10^{-3}$ and $2 \times 10^{-4}$, respectively, in reasonable agreement with the numerically simulated values of $\alpha$. Thus, Eqs. (7-8) with $\alpha$ calculated analytically, as obtained from the simplified electrical model shown in Extended Data Fig. 9b, also allow for semi-quantitative estimates of $\sigma_{//}$ and $\varepsilon_{//}$, which are found to differ only by a factor of < 2 from the values extracted through our more quantitative, numerical analysis.

**Acknowledgements** We thank A. Michaelides, C. Schran and J. Chen for useful discussions; I. Horcas and J. Gómez-Herrero for support with AFM. This research was funded by the European Research Council (grants 819417-Liquid2DM and VANDER), the Marie Sklodowska-Curie program (grants 793394, 842402 and 873028), the Leverhulme Trust (grant RPG-2023-253), the UKRI (grant EP/X022471/1) and the Lloyd's Register Foundation.

**Author contributions** L.F. designed and directed the research with help from A.K.G. R.W., M.S. and A.E. fabricated most of the devices with contributions of Q.Y., H.N-A, and J.N. S.B., H.N-A, P.A. and L.F. carried out experiments and analyzed the data. R.F. implemented finite-element numerical calculations. P.A. provided AFM instrumentation. G.F. provided electronic instrumentation. A.P. provided point-charge theoretical calculations. L.F. developed the electrical model with help of G.F. A.K.G. and L.F. wrote the manuscript with contributions of A.P. All authors contributed to discussions.

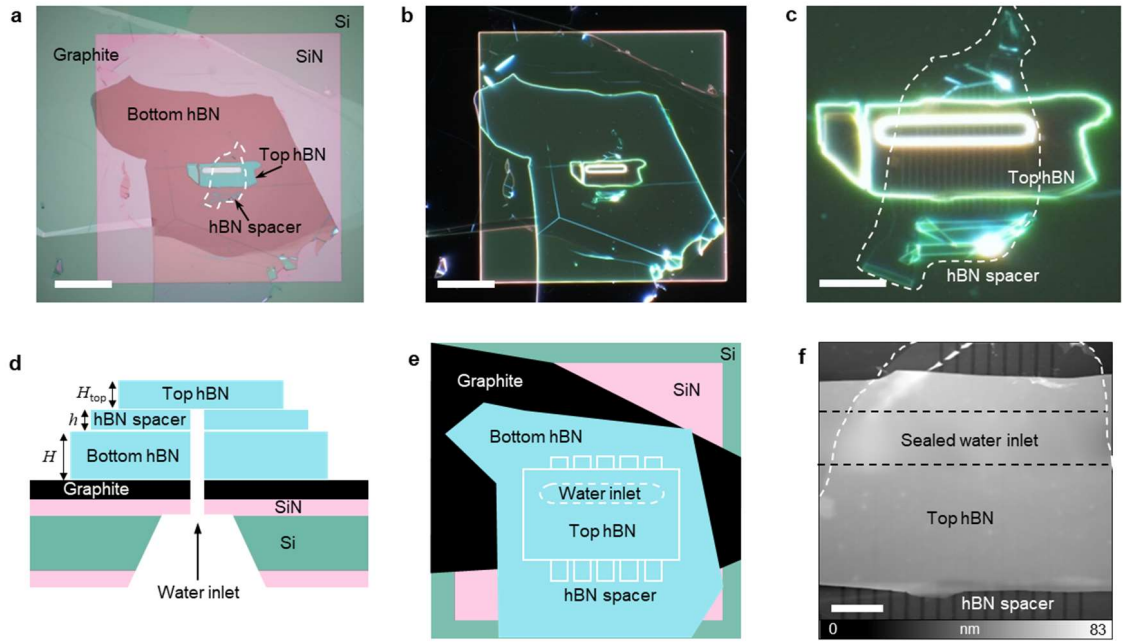

**Extended Data Fig. 1 | Devices for in-plane dielectric imaging of confined water. (a)** Optical image of a representative device with the channel height $h \approx 5$ nm. The edges of the spacer layer are indicated by the white dashed lines. The large square region (appears in pink) is the suspended SiN membrane. It has a ~ 20 μm-long rectangular slit in its center (white) that connects nanochannels to a water reservoir at the backside of the device. **(b)** Dark-field optical image of the same device and **(c)** Zoom into its central region where numerous parallel channels can be seen. **(d,e)** Sketch of the device geometry. Its cross-section (d) and top-view (e) (not to scale). **(f)** AFM topography image of the device's central region. The black dashed lines indicate the aperture's position. The white dashed curve indicates edges of the hBN spacer crystal. Scalebars: (a,b) 20 μm; (c) 10 μm; and (f) 3 μm.

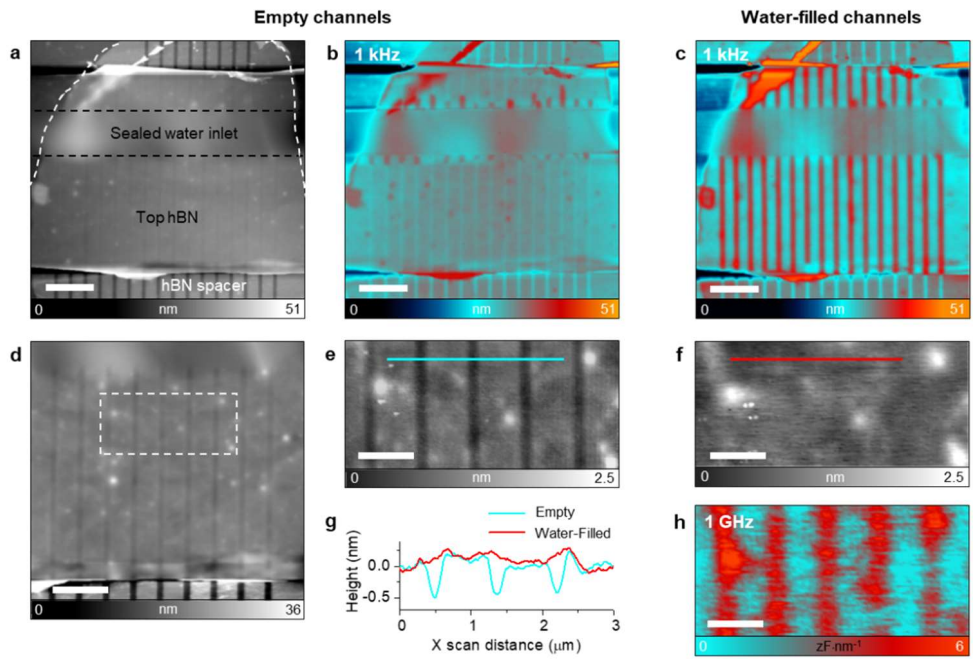

**Extended Data Fig. 2| Filling nanochannels with water. (a)** AFM topography image of a representative device with $h \approx 5$ nm before filling it with water – same image as in Extended Data Fig. 1f but using a flattening filter to better visualize the sagging of the top hBN layer into nanochannels. The image was taken in the intermittent-contact attractive mode with no voltage applied. **(b,c)** Same as (a) but with applying a low-$f$ electric field ($v_{AC}$ = 8V, 1 kHz) before (b) and after (c) filling the nanochannels with water. The contrast associated with the channels changes from light cyan in (b) to red in (c), indicating that water filled all the channels. **(d)** Zoom-in of the central region in (a). The top hBN layer is partially sagged into the empty channels. **(e)** Further zoom-in showing the region indicated by the dashed rectangular in (d). **(f)** Same as (e) but after filling nanochannels with water. It shows that the top hBN layer became flat, no longer sagging inside the nanochannels. **(g)** Topography profiles over three channels in (e,f), as indicated by color-coded lines. **(h)** Dielectric image taken from the same region as in (f) at a constant height and at 1 GHz. Scalebars: (a-c) 3 µm, (d) 2 µm, (e,f,h) 1 µm.

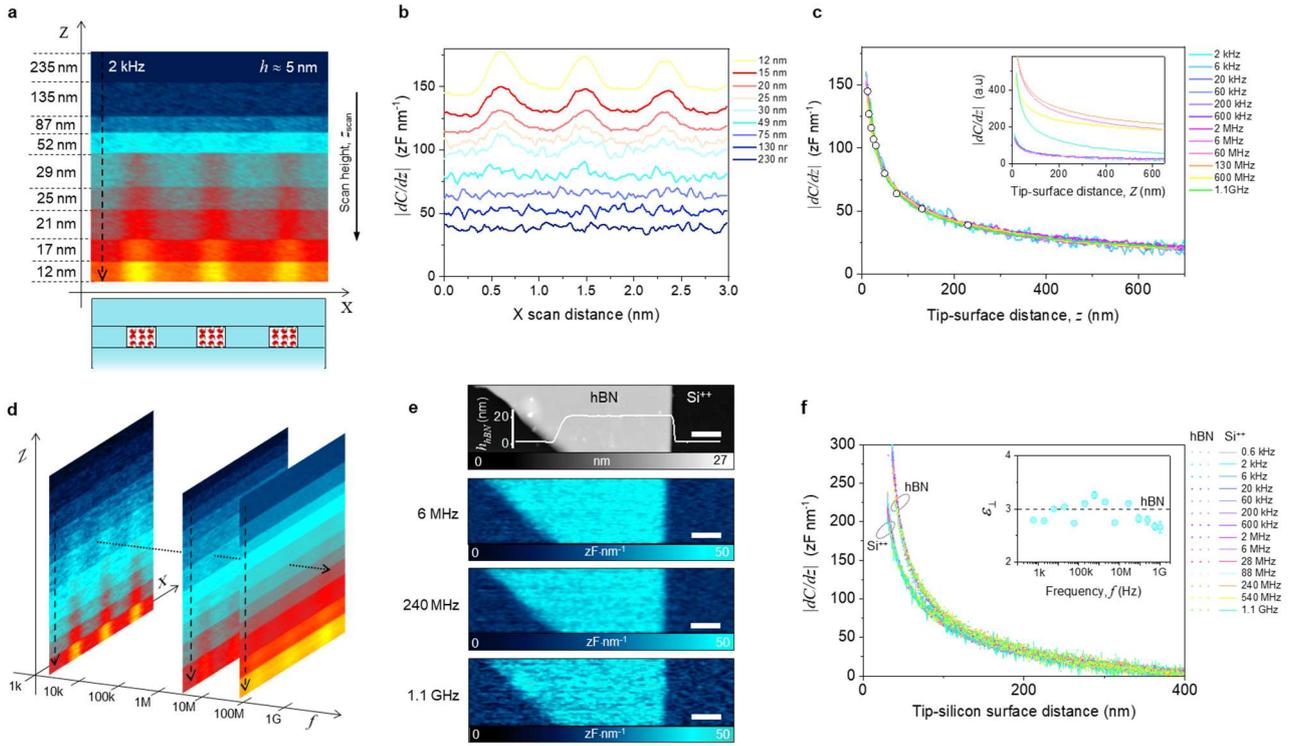

**Extended Data Fig. 3 | Measurement and calibration procedures for local broadband dielectric spectroscopy. (a)** Representative $|dC/dz|(x,z)$ image taken at 2 kHz for channels with $h \approx 5$ nm. The image shows 3 nanochannels and was acquired by measuring $|dC/dz|$ at various constant heights and a fixed $y$ position. The tip approached the surface from above using small steps. **(b)** $|dC/dz|$ for scanning along the lateral ($x$) axis at constant $y$ and various heights. The profiles were averaged over ~10 lines using images in (a). The signal became sensitive to the presence of water inside the channels at $z \lesssim 30$ nm. **(c)** Gain-calibrated $|dC/dz|(z)$ curves at different frequencies taken over the center of the spacer region (color-coded lines). White symbols indicate the values of $|dC/dz|$ and $z_{scan}$ corresponding to the profiles shown in (b). Inset: Same curves without calibration of the electrical gain $g$ and the offsets (color-coded). We used the curve at 2 kHz as a reference to calibrate $g$ at high frequencies by matching the reference curve with the curves taken at higher $f$, using only $g$ and an offset as fitting parameters. **(d)** Schematic representation of our 3D AFM data sets $|dC/dz|(x,z,f)$ obtained by acquiring subsequent images $|dC/dz|(x,z)$ for many frequencies from kHz to GHz. The black dashed arrows in (a) and (d) indicate the position where the $|dC/dz|(z)$ curves in (c) were acquired before taking such 3D images. **(e,f)** Example of broadband dielectric imaging spectroscopy on a test sample of hBN on doped silicon using the same $g$ calibration procedure: topography image (grey colored) of 21 nm-thick hBN crystal on doped Si; and corresponding calibrated dielectric images (cyan colored images) at frequencies 6 MHz, 240 MHz and 1.1 GHz at a constant height (~45 nm). Scale bar, 1 µm. The curves in (f) are gain-calibrated $|dC/dz|(z)$ curves taken above the Si substrate (solid curves) and the hBN layer (dotted) at various $f$. The gain $g$ for each $f$ was obtained by scaling the curves measured on doped Si to the low-$f$ reference curve at 2 kHz using only $g$ and an offset as fitting parameters. By fitting the calibrated curves on hBN using the procedure and analytical formula in Ref.[65], we obtained the dielectric constant of the hBN layer as function of $f$ (inset). Its average value across all the $f$ up to GHz was found to be $\varepsilon_{\perp hBN} = 2.9 \pm 0.2$, which is in good agreement with the expected bulk value for hBN in the perpendicular direction ($\varepsilon_{\perp hBN} \approx 3$, dashed line). Bars: standard error defined as 95% confidence interval of the fitting, not shown when smaller than the symbol. Other fitting parameters: AFM tip's $R = 102$ nm and $\theta = 25°$, obtained by fitting the curve taken on doped Si at 2 kHz.

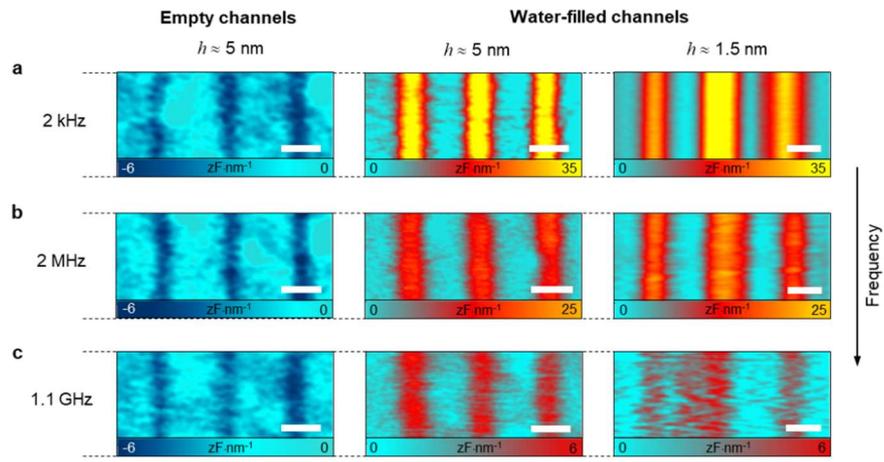

**Extended Data Fig. 4| Additional dielectric images.** (**a-c**) Dielectric images of three channels taken at constant height and at characteristic $f$ of 2 kHz, 2 MHz and 1.1 GHz, respectively. Same devices as in Fig. 2b of the main text: $h \approx 5$ and 1.5 nm, empty and filled with water (as specified in the legends). Scalebars, 500 nm.

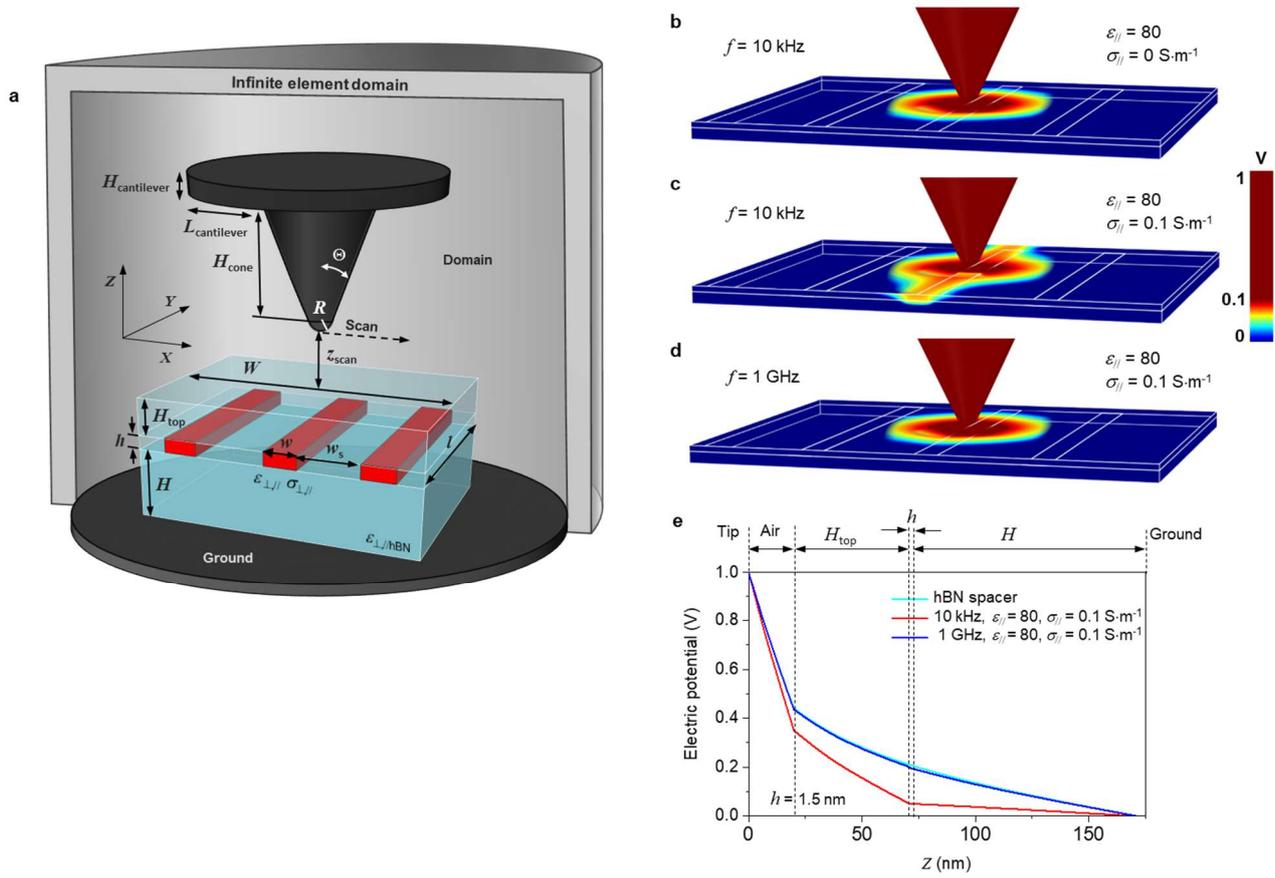

**Extended Data Fig. 5 | 3D numerical simulations.** (**a**) Schematics (not to scale) of the 3D numerical model used to fit the experimental data. (**b**-**d**) Calculated electrostatic potential for $h$ = 1.5 nm and three representative cases: (b) low (10 kHz) frequency and purely dielectric channels ($\varepsilon_{//}$ = 80 and $\sigma_{//}$ = 0); (c) Same low frequency but conductive channels ($\varepsilon_{//}$ = 80 and $\sigma_{//}$ = 0.1 S·m$^{-1}$); (d) same as (c) but at $f$ beyond conductivity relaxation. (**e**) Potential along the $Z$ direction (counted from the AFM tip) for the tip placed at $z_{scan}$ = 20 nm above the middle of the water channel (red and blue) and above the center of the spacer region (cyan). Frequencies 10 kHz and 1 GHz, color coded. Water channel's parameters: $\varepsilon_{//}$ = 80, $\sigma_{//}$ = 0.1 S·m$^{-1}$, $\varepsilon_\perp$ = 2, $\sigma_\perp$ = 0, $l$ = 3 µm, $w_c$ = 200 nm; $w_s$ = 800 nm. Other parameters: $R$ = 100 nm, $\theta$ = 25°, $H_{cone}$ = 6 µm, $H_{cantilever}$ = 3 µm, $L_{cantilever}$ = 0 µm, $H_{top}$ = 50 nm, $H$ = 100 nm, $\varepsilon_{//hBN}$ = 5, $\varepsilon_{\perp hBN}$ = 3.

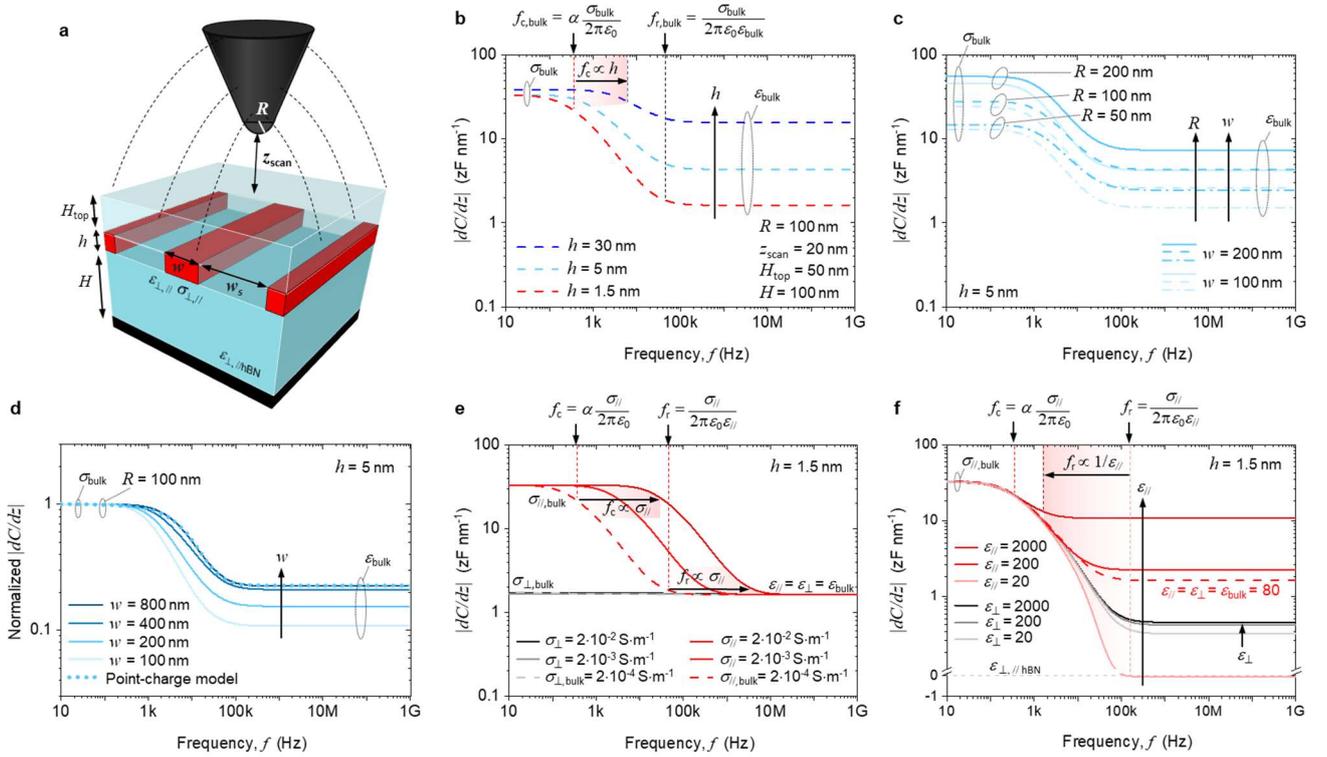

**Extended Data Fig. 6 | Calculated dielectric spectra using the 3D numerical simulations.** (**a**) Simplified illustration of our numerical model (shown in full in Extended Data Fig. 5). (**b**) Calculated $|dC/dz|$ as a function of frequency for different channel thicknesses ($h$ = 30, 5 and 1.5 nm) and bulk water ($\varepsilon_{//} = \varepsilon_{\perp}$ = $\varepsilon_{bulk}$, $\sigma_{//} = \sigma_{\perp} = \sigma_{bulk}$) inside the channels. (**c**) Same as in (b) for $h$ = 5 nm but using various AFM tip radii and different channel widths. (**d**) Same as in (b,c) but for several $w$; normalized by the height of the low-$f$ plateau. Dotted curve: corresponding calculations using our point-charge analytical model. (**e**) Same as (b) but for $h$ = 1.5 nm and different water's $\sigma_{//}$ (red-colored curves, keeping $\sigma_{\perp}$ = 0,) and different $\sigma_{\perp}$ (grey-colored curves, $\sigma_{//}$ = 0). (**f**) Same as in (e) but for different $\varepsilon_{//}$ (red-colored curves, keeping $\varepsilon_{\perp}$ = 2) and different $\varepsilon_{\perp}$ (grey-colored curves, $\varepsilon_{//}$ = 2). The red dashed curve is the case of the channel filled with bulk water, same as shown in (b). Other parameters used in (b-f) (unless stated otherwise in the legends): $R$ = 100 nm, $\theta$ = 25°, $H_{cone}$ = 6 μm, $H_{cantilever}$ = 3 μm, $L_{cantilever}$ = 0 μm, $z_{scan}$ = 20 nm; $l$ = 3 μm, $w_c$ = 200 nm, $w_s$ = 800 nm, $H_{top}$ = 50 nm and $H$ = 100 nm. All the plotted $|dC/dz|$ curves are after subtracting the $|dC/dz|$ values found at the center of the hBN spacer region.

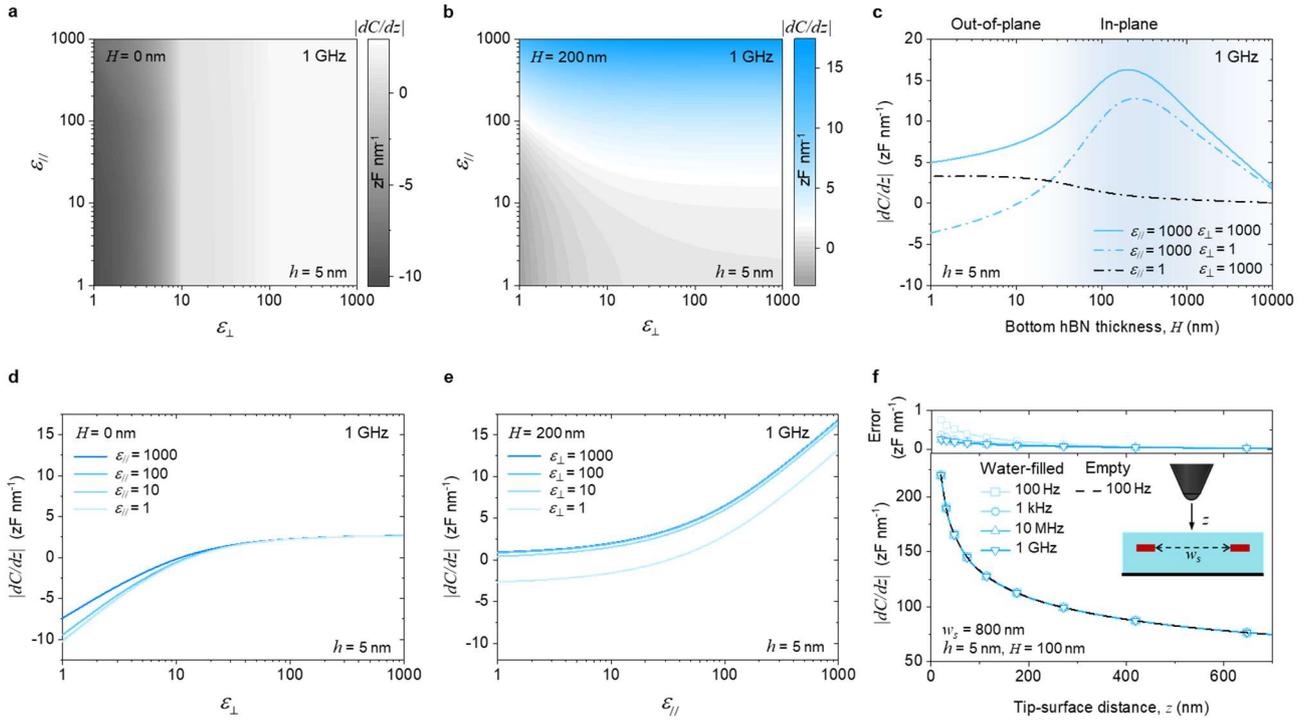

**Extended Data Fig. 7 | Sensitivity to water's in-plane and out-of-plane dielectric constants and calibration curves using the 3D numerical simulations.** (**a,b**) Calculated $|dC/dz|$ at 1 GHz as a function of water's $\varepsilon_{//}$ and $\varepsilon_{\perp}$ (with $\sigma_{//} = \sigma_{\perp} = \sigma_{bulk}$). $h$ = 5 nm and the bottom hBN thicknesses (a) $H$ = 0 (metallic substrate) and (b) $H$ = 200 nm. (**c**) Calculated $|dC/dz|$ as in (a) and (b) but as a function of $H$ for different combinations of $\varepsilon_{//}$ and $\varepsilon_{\perp}$. AFM cone height $H_{cone}$ = 12 μm and cantilever length $L_{cantilever}$ = 20 μm (nominal parameters) to include long-range range effects that play a role for very thick $H$. (**d**) Cross-sections from (a) as a function of water's $\varepsilon_{\perp}$ for different values of $\varepsilon_{//}$. (**e**) Cross-sections from (b) as a function of $\varepsilon_{//}$ for different $\varepsilon_{\perp}$. (**f**) Calculated $|dC/dz|$ vs tip-surface distance above the center of the hBN spacer, as illustrated in the inset, for empty channels (black dashed line) and bulk-water-filled channels (cyan colored symbols) at various frequencies. The channel thickness is $h$ = 5 nm and the bottom hBN thickness is $H$ = 100 nm. Upper panel: error between the calculated curves for water-filled channels and that for empty channels. Other parameters used in (a-f) (unless stated otherwise in the legends): $R$ = 100 nm, θ = 25°, $H_{cone}$ = 6 μm, $H_{cantilever}$ = 3 μm, $L_{cantilever}$ = 0 μm, $z_{scan}$ = 20 nm; $l$ = 3 μm, $w_c$ = 200 nm, $w_s$ = 800 nm, $H_{top}$ = 50 nm. All the plotted $|dC/dz|$ curves in panels (a-e) are after subtracting the $|dC/dz|$ values found at the center of the hBN spacer region.

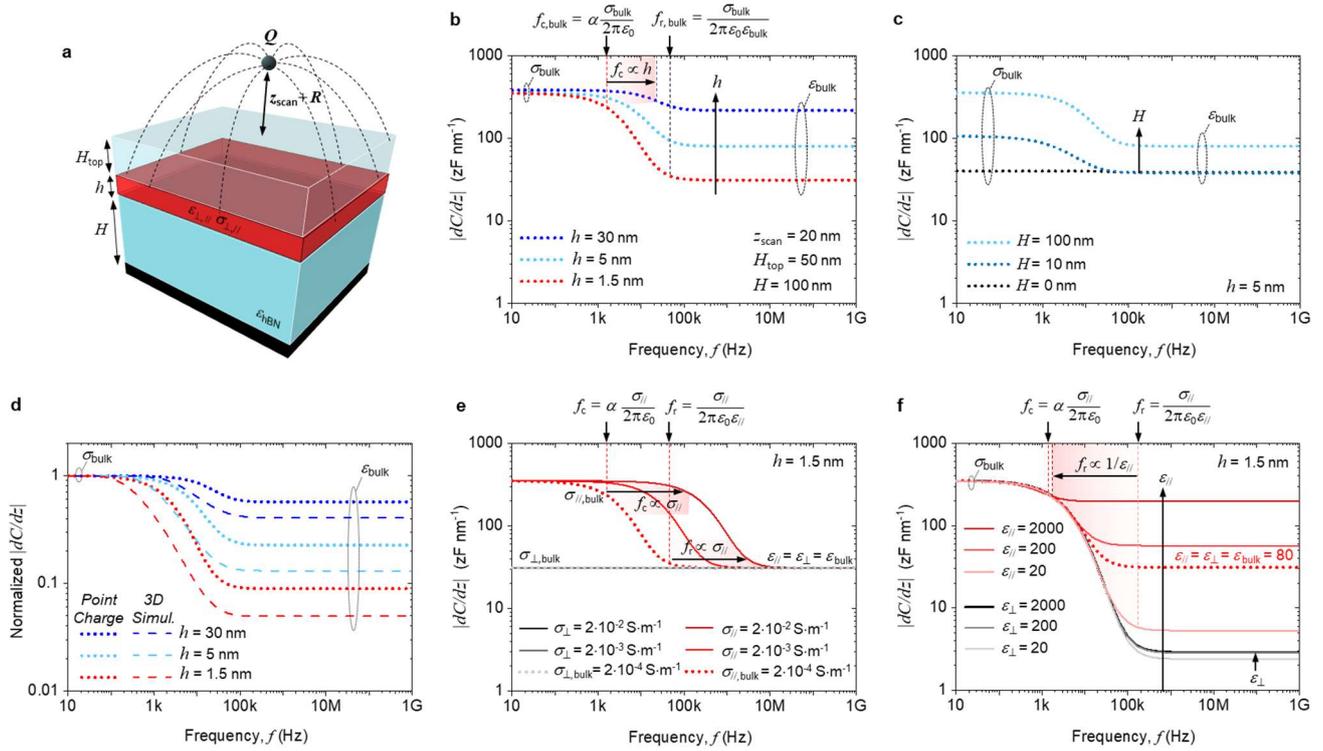

**Extended Data Fig. 8 | Calculated dielectric spectra using the point-charge analytical model.** (**a**) Its schematic. A point charge is used to approximate the AFM tip. The device is modeled as a water layer of infinite lateral size (red), which is encapsulated between top and bottom hBN layers (blue). For simplicity, the dielectric constant of hBN is assumed to be isotropic and equal to 4. (**b**) Calculated $|dC/dz|$ as a function of frequency for $h$ = 30, 5 and 1.5 nm (color coded) and bulk water ($\varepsilon_{//} = \varepsilon_\perp = \varepsilon_{bulk}$, $\sigma_{//} = \sigma_\perp = \sigma_{bulk}$). (**c**) Same as in (b) for $h$ = 5 nm but using various thicknesses of bottom hBN. (**d**) The dotted curves are the same as in (b) but normalized with respect to the low-$f$ plateau's height. Dashed curves: corresponding analysis using the full 3D numerical simulations. (**e**) Same as in (b) for $h$ = 1.5 nm and different $\sigma_{//}$ (red curves, keeping $\sigma_\perp$ = 0) and $\sigma_\perp$ (grey, $\sigma_{//}$ = 0). (**f**) Same as in (b) for $h$ = 1.5 nm and various $\varepsilon_{//}$ (red solid curves, keeping $\varepsilon_\perp$ = 2) and various $\varepsilon_\perp$ (grey curves, $\varepsilon_{//}$ = 2). The red dotted curve is for the same channel filled with bulk water [also shown in (b)]. Other parameters used in (b-f) (unless stated otherwise in the legends) are $z_{scan}$ = 20 nm, $H_{top}$ = 50 nm and $H$ = 100 nm. All the plotted $|dC/dz|$ curves are after subtracting the $|dC/dz|$ values found for the same geometry but with hBN instead of the water layer.

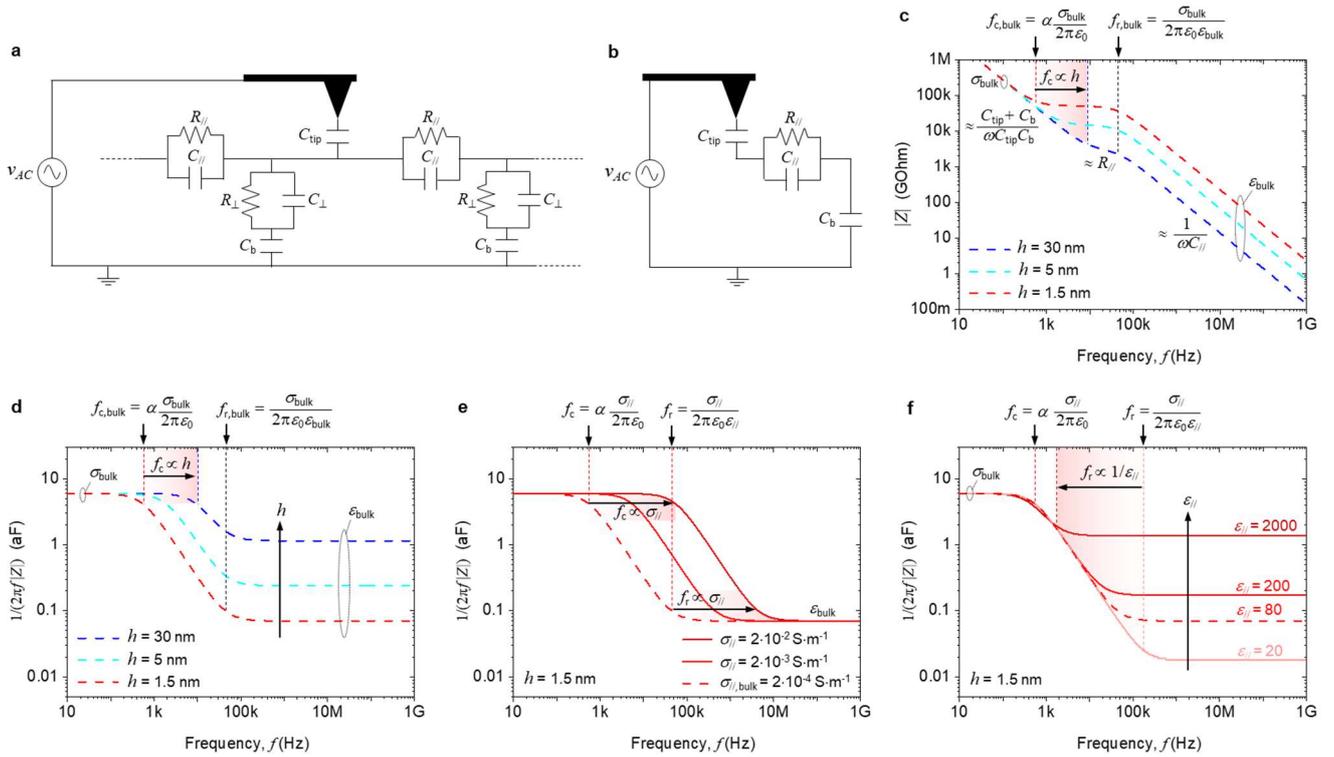

**Extended Data Fig. 9 | Simulated dielectric spectra using equivalent electrical circuits.** (**a**) Approximate impedance model for our measurement geometry using a distributed RC network that includes both in-plane and out-of-plane elements to model the water channel. (**b**) Simplified in-plane equivalent circuit that takes into account the negligible contribution of the out-of-plane impedance. (**c**) Calculated impedance using different channel thicknesses ($h$ = 30, 5 and 1.5 nm) and bulk water's properties. (**d**) Same as in (c) but plotting the effective capacitance, $1/(|Z|2\pi f)$. (**e**) Same as (d) for $h$ = 1.5 nm using different $\sigma_{//}$. (**f**) Same as (e) but for different $\varepsilon_{//}$. Other parameters used in simulations (c-f): $R$ = 100 nm, $\theta$ = 25°, $z_{scan}$ = 20 nm, $l^*$ = 3 μm, $w_c$ = 200 nm, $H_{top}$ = 50 nm and $H$ = 100 nm.

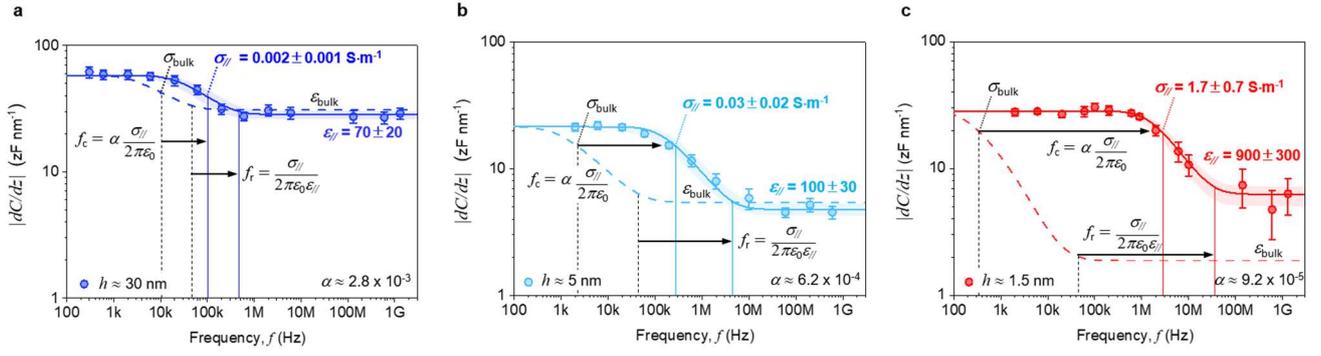

**Extended Data Fig. 10| Fitting experimental spectra. (a-c)** Symbols are the experimental data of Fig. 2a of the main text before normalization. Bars are the random noise level as measured at each $f$. The solid curves are the best fits using the 3D numerical model with water's $\varepsilon_{//}$ and $\sigma_{//}$ as the only fitting parameters (labels). The light-shaded regions indicate the uncertainty in the extracted values of $\varepsilon_{//}$ and $\sigma_{//}$. Dashed curves: 3D numerical simulations for the same devices assuming bulk water's properties ($\varepsilon_{//} = \varepsilon_{\perp} = \varepsilon_{bulk} = 80$, $\sigma_{//} = \sigma_{\perp} = \sigma_{bulk} = 2 \times 10^{-4}$ S·m$^{-1}$). The vertical lines indicate calculated values of $f_c$ and $f_r$ for the case of bulk water (black dashed) and the experimentally extracted water parameters (colored solid). To evaluate the frequencies, we used Eqs. (5-6) and $\alpha$ obtained from the full 3D numerical simulations: **(a)** $\alpha = 2.8 \times 10^{-3}$, **(b)** $\alpha = 6.2 \times 10^{-4}$ and **(c)** $\alpha = 9.2 \times 10^{-5}$. $\varepsilon_{\perp}$ is set to the values measured in Ref.[22]. Other experimental parameters used in the simulations: **(a)** $h$ = 30 nm, $\varepsilon_{\perp}$ = 32, $\sigma_{\perp}$ = 2·10$^{-4}$ S·m$^{-1}$, $R$ = 84 nm, $\theta$ = 26°, $H_{top}$ = 21 nm, $H$ = 53 nm, $w$ = 210 nm, $z_{scan}$ = 19 nm; **(b)** $h$ = 5 nm, $\varepsilon_{\perp}$ = 6.5, $\sigma_{\perp}$ = 2·10$^{-4}$ S·m$^{-1}$, $R$ = 62 nm, $\theta$ = 17°, $H_{top}$ = 34 nm, $H$ = 45 nm; $w$ = 220 nm, $z_{scan}$ = 18 nm; **(c)** $h$ = 1.5 nm, $\varepsilon_{\perp}$ = 2.1, $\sigma_{\perp}$ = 2·10$^{-4}$ S·m$^{-1}$, $R$ = 91 nm, $\theta$ = 27°, $H_{top}$ = 50 nm, $H$ = 76 nm, $w$ = 210 nm, $z_{scan}$ = 19 nm.